\documentclass[conference]{IEEEtran}
\IEEEoverridecommandlockouts

\usepackage{cite}
\usepackage{amsmath,amssymb,amsfonts}
\usepackage{tikz}
\usetikzlibrary{shapes.arrows} 
\usepackage{amsmath}
\usepackage{filecontents}
\usepackage{lipsum}
\usepackage{tikz}
\usepackage{kotex}
\usepackage{tabularx}
\usepackage{algpseudocode}
\usepackage{graphicx}
\usepackage{setspace}
\usepackage{textcomp}
\usepackage{xcolor}
\usepackage{caption}
\usepackage{subcaption}
\usepackage{color}
\usepackage{float}
\usepackage{comment}
\usepackage[normalem]{ulem}
\usepackage{titlecaps}
\usepackage{wrapfig}
\usepackage{multirow}
\usepackage{array}
\usepackage{pifont}
\usepackage{mathtools}
               
\usepackage{xurl}  

\newcommand{\tbdfixed}{{\textsc{DiaLSM}}}

\usepackage{lipsum}
\usepackage{enumitem}
\usepackage{url}
\usepackage{hyperref}
\hypersetup{hidelinks,colorlinks=true,linkcolor=green!80!black,citecolor=red!70!black,urlcolor=blue!70!black}
\usepackage{bm}
\usepackage{colortbl}

\usepackage{tikz}
\newcommand*\circled[1]{\tikz[baseline=(char.base)]{
    \node[shape=circle,draw,fill=white,text=black,inner sep=1pt, line width=1pt] (char) {\bfseries #1};}}

\usepackage[linesnumbered,ruled,vlined]{algorithm2e}

\SetCommentSty{mycommfont}
\SetKwInput{KwInput}{Input}                
\SetKwInput{KwOutput}{Output}              

\newcommand{\squishlist} {
\begin{list}{$\bullet$}
	{ \setlength{\itemsep}{0pt}      
       \setlength{\parsep}{-0pt}
		\setlength{\topsep}{4pt}      \setlength{\leftmargin}{0pt}
        \setlength{\itemindent}{-1em}
        \setlength{\partopsep}{0pt}
		\setlength{\listparindent}{-2pt}
		\setlength{\itemindent}{-5pt}
		\setlength{\leftmargin}{1em} 
        \setlength{\labelwidth}{0em}
		\setlength{\labelsep}{0.5em} } }
\newcommand{\squishend}{
\end{list}  }

\newcommand{\subsquishlist}{
\begin{list}{-}
  { \setlength{\itemsep}{0pt}      
    \setlength{\parsep}{0pt}
    \setlength{\topsep}{2pt}       
    \setlength{\partopsep}{0pt}
    \setlength{\listparindent}{-2pt}
    \setlength{\itemindent}{-5pt}
    \setlength{\leftmargin}{1em} 
    \setlength{\labelwidth}{0em}
    \setlength{\labelsep}{0.5em} } }

\newcommand{\subsquishend}{
\end{list}}

\usepackage{hhline} 
\usepackage{xfp}
\usepackage{adjustbox}
\usepackage{booktabs}
\usepackage{orcidlink}

\begin{document}

\title{\tbdfixed{}: Towards Write-Stall-Free Performance via Shard-based LSM-tree
\vspace{-10pt}
}

\setlength{\skip\footins}{6pt}

\author{\IEEEauthorblockN{Hongsu Byun\textsuperscript{1,}\IEEEauthorrefmark{1}\orcidlink{0000-0002-2143-4292}, Safdar Jamil\textsuperscript{3}\orcidlink{0000-0002-9011-6431}, Honghyeon Yoo\textsuperscript{2}\orcidlink{0009-0008-4935-3970},\\Sungyong Park\textsuperscript{2}\orcidlink{0000-0002-0309-1820}, Myungcheol Lee\textsuperscript{4}\orcidlink{0000-0002-1251-1727}, Xubin He\textsuperscript{5}\orcidlink{0000-0002-5071-2861}, Zhichao Cao\textsuperscript{6}\orcidlink{0000-0001-6950-1776}, Youngjae Kim\textsuperscript{2,}\IEEEauthorrefmark{2}\orcidlink{0000-0001-8786-3850}
\thanks{\IEEEauthorrefmark{1}This work was conducted while the author was at Sogang University.
}
\thanks{\IEEEauthorrefmark{2}Y. Kim is the corresponding author.
}
}

\IEEEauthorblockA{
\textsuperscript{1}\textit{Korea Aerospace University},
\textsuperscript{2}\textit{Sogang University},
\textsuperscript{3}\textit{MangoBoost},
\textsuperscript{4}\textit{ETRI},
\textsuperscript{5}\textit{Temple University},
\textsuperscript{6}\textit{Arizona State University}\\
hsbyun@kau.ac.kr, \{yhh, parksy, youkim\}@sogang.ac.kr, safdar.jamil@mangoboost.io, mclee@etri.re.kr,\\xubin.he@temple.edu, zhichao.cao@asu.edu
}
\vspace{-25pt}
}

\maketitle

\begin{abstract}
Log-Structured Merge-tree (LSM) aims to achieve high write throughput, but is known to experience the write stall problems when subjected to sustained write pressure. We quantify the occurrence probability and average duration of write stalls in LSM using a queuing model in the write--flush--compaction pipeline, moving beyond existing empirical analysis. The proposed model demonstrates that a monolithic LSM with a single pipeline cannot eliminate write stalls, revealing that internal sharding within the LSM offers an opportunity for fundamental write stall mitigation.

To break this structural bottleneck, we propose \tbdfixed{}, an internally shard-based LSM architecture. Instead of forcing all writes through one pipeline, \tbdfixed{} splits the write--flush--compaction path into multiple independent shards and employs dynamic fallback redirection, allowing writes to proceed even when some shards stall.
Implemented on RocksDB, \tbdfixed{} achieves up to 2.4$\times$ higher throughput, 94\% lower stalls, and significantly lower latency than state-of-the-art methods ADOC and Sub-Compaction, as demonstrated by db\_bench, YCSB, and Sysbench OLTP evaluations. 
\end{abstract}


\maketitle

\setlength{\textfloatsep}{5pt plus 1.0pt minus 2.0pt}
\vspace{-6pt}
\section{Introduction}
\label{sec:intro}
\noindent\textbf{Monolithic LSM tree.}
Log-Structured Merge-tree (LSM tree)~\cite{o1996log} is a write-optimized data structure widely used in NoSQL systems such as LevelDB~\cite{levelDB}, RocksDB~\cite{rocksdb}, and Cassandra~\cite{cassandra}. 
Although LSM trees are often described by data layout and compaction policies, they form a tightly coupled \emph{data pipeline}. Incoming writes are first buffered in a Memtable, flushed as immutable SSTables to Level~0 (L0), and incrementally compacted into lower levels (L$1$–L$d$) to maintain key order and control storage amplification. Crucially, these stages—write buffering, flushing, and compaction—are not independent. As illustrated in Figure~\ref{fig:back_lsm}, they are organized as a \emph{single, monolithic pipeline}, where progress at each stage directly depends on the throughput of downstream stages. This architectural coupling, rather than any specific compaction policy or parameter choice, forms the structural foundation on which subsequent performance limitations arise.

\begingroup
\setlength{\textfloatsep}{0pt}  
\begin{figure}[!t]
	\centering
    \includegraphics[width=0.95\linewidth]{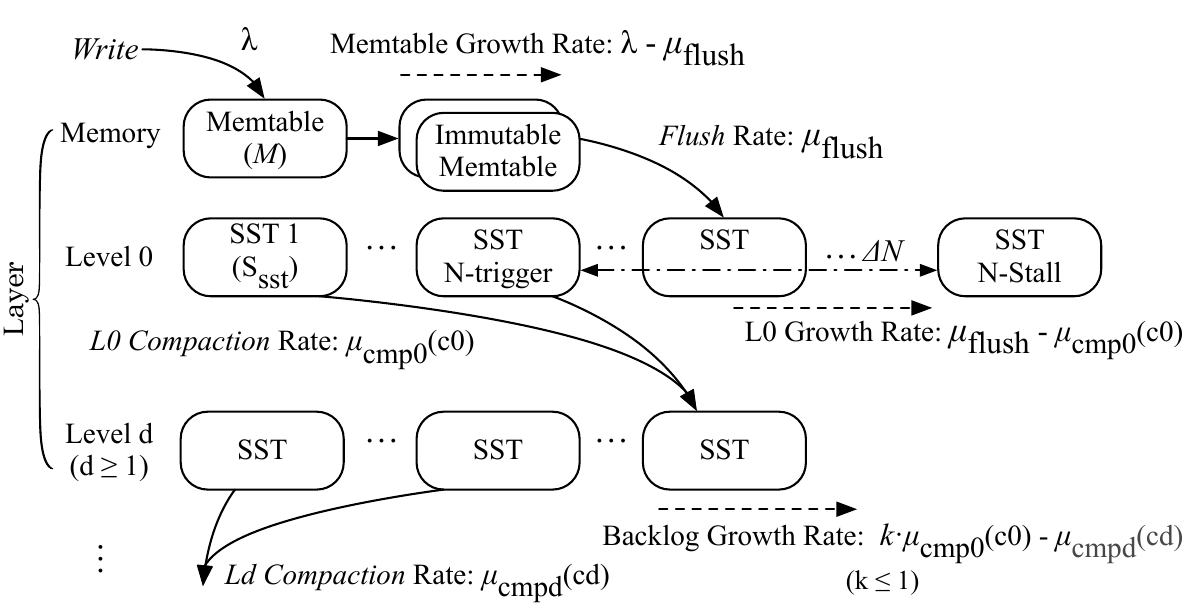}
    \vspace{-6pt}
    \caption{\small Description of the data pipeline in a monolithic LSM tree architecture. Layer refers to the Memtable and levels.
	}
    \vspace{-5pt}
	\label{fig:back_lsm}
\end{figure}
\endgroup

\noindent\textbf{Write Stall.}
The core advantage of the LSM tree is that it provides stable write performance even under frequent random writes by relying on sequential writes and deferred sorting through background compaction. However, ironically, once the write load exceeds a certain threshold, the LSM tree can no longer accept writes and instead triggers a \textit{write stall}~\cite{yu2023adoc, yao2020matrixkv, balmau2019silk, ding2022trianglekv, he2023flatlsm, luo2019performance, liang2021cruisedb} that blocks write requests. In this state, foreground write I/O is blocked until internal flush or compaction operations make progress, temporarily preventing the system from responding to external requests and causing high latency. Depending on which layer of the internal pipeline, such as the Memtable, Level~0, or deeper levels, becomes the bottleneck, different types of stalls occur, leading paradoxically to a situation in which a write-optimized LSM tree halts its own writes under heavy workloads.

\noindent\textbf{Limitations of the Monolithic LSM tree.} 
Existing solutions have proposed various techniques to mitigate write stalls. For example, ADOC~\cite{yu2023adoc} adjusts the Memtable size and the number of compaction threads, while Sub-Compaction~\cite{subcompaction} parallelizes L0-to-L1 compaction to better reduce compaction pressure at L0. Although these approaches reduce the frequency or duration of write stalls at specific layers, they share a fundamental limitation in that they operate within the structural constraint of a single LSM pipeline. That is, while they can change where and to what extent stalls occur, they cannot eliminate write stalls entirely.

\noindent\textbf{Modeling for Write Stall.}
Write stalls are not a problem of a suboptimal implementation or tuning heuristics, but a phenomenon that appears across the monolithic LSM tree architecture. However, empirical analysis~\cite{yu2023adoc,luo2019performance} alone is insufficient to explain why write stall persists and why it cannot be eliminated under any conditions. To systematically understand this behavior, we introduce a model that enables mathematical analysis of the LSM tree write path. 

We view the LSM tree write path, consisting of write--flush--compaction, as a single execution pipeline, and model situations in which foreground write processing is temporarily suspended due to flush or compaction using the notion of vacations in queueing theory~\cite{fuhrmann1985stochastic}.
\textit{The goal of this model is not parameter tuning or performance prediction, but to show that write stalls are structurally inevitable as long as the single LSM pipeline structure is maintained.} The detailed modeling in Section~\ref{sec:model} explains why write stalls cannot be eliminated in monolithic LSM trees and derives why a new design direction is required.

\noindent\textbf{Sharding as a Structural Solution.} To overcome limitation of monolithic LSM tree, we revisit sharding~\cite{abdelhafiz2021sharding, xiao2015shardfs, kvrocks, zippydb}---a proven scalability technique in distributed databases. By applying sharding within a single LSM instance, we decompose the write--flush--compaction pipeline into multiple independent pipelines, dramatically reducing stall probability and improving robustness. This allows write traffic to be load-balanced and redirected across shards, ensuring progress even if individual shards stall.

\noindent\textbf{\tbdfixed{}.} 
In this paper, we present \tbdfixed{} (Diamond-LSM), a shard-based LSM architecture that structurally departs from the monolithic model. Like a diamond with many facets, \tbdfixed{} partitions a single logical tree into multiple internal LSM shards. It significantly reduces write stalls in practice through two mechanisms: hash-based load balancing and the redirection of stalled writes to fallback shards. While theoretical corner cases exist where all shards might stall, in practice, this is extremely rare, making stalls negligible in frequency compared to monolithic designs. 

\noindent\textbf{Challenges.} With this structural shift, several core aspects that are naturally supported in monolithic LSM-trees require rethinking and redesign in a shard-based architecture, leading us to address the following critical challenges:

\squishlist
\item \textit{Indexing for Fallback Shards.}  Redirected writes complicate lookup processing, as keys may reside outside their primary shards.  A mechanism is required for looking up keys located outside the primary shard.

\item \textit{Ensuring Read Latency.} Lookups across primary and fallback shards can add extra latency. Read performance must also be guaranteed at the level of traditional LSM-tree.

\item \textit{Preserving Ordering.} Hash-based sharding disrupts key locality, making it challenging to support ordered access patterns like range queries.

\item \textit{Providing Robustness.} When a system crash occurs during redirection writing, data consistency may be broken. Failure recovery and consistency maintenance must be ensured.
\squishend

\noindent\textbf{Contributions.} This paper makes the following contributions.

\squishlist
\item 
We quantify write stall occurrence based on a queueing model in a monolithic LSM-tree with a single pipeline structure (§\ref{sec:model}).

\item 
We demonstrate that existing techniques induce cross-layer balloon effects, preventing write stall elimination (§\ref{sec:motivation}), and highlight the need for structural alternatives to fundamentally mitigate write stalls (§\ref{sec:opportunity}).

\item 
We propose \tbdfixed{}, a shard-based LSM design that significantly reduces write stalls through dynamic fallback writes, while introducing Clue Entries (CEs), lightweight pointers stored in primary shards to track redirected keys even during stall conditions (§\ref{sec:design}).

\item 
We design a set of mechanisms that enable shard-based LSM by preserving global ordering, efficient range queries, and consistency, including CE L0 capping and CE caching to reduce lookup latency, parallel iterators without extra storage, and garbage collection and rollback policies to restore data to primary shards.
\squishend

We implemented \tbdfixed{} atop RocksDB v8.3.2 and evaluated it with db\_bench and YCSB benchmark. In the FillRandom workload, \tbdfixed{} reduced write stalls by 94.3\% and achieved 2.4$\times$ higher throughput and 50\% lower latency than ADOC. In the ReadRandomWriteRandom workload, it delivered 1.8$\times$ higher throughput. Across all YCSB workloads, \tbdfixed{} outperformed ADOC, with up to 8.9$\times$ gains on YCSB-E due to improved range query performance. 
Furthermore, in Sysbench-based OLTP evaluations on RocksDB-based RDBMSs, \tbdfixed{} achieves 2.2$\times$ higher throughput than ADOC under mixed read/write workloads.
\section{Modeling Write Stalls in LSM trees}
Empirical analysis~\cite{yu2023adoc,balmau2019silk,yao2020matrixkv,luo2019performance} cannot fully explain the underlying dynamics of LSM tree write stalls, as the interactions among buffering, flushing, and compaction are tightly coupled. To capture these dynamics systematically, 
we adopt a mathematical analysis model. This abstraction allows us to classify stall types, derive their occurrence characteristics, and establish closed-form expressions for stall time --- laying the foundation for analysis and stall-mitigation design. The modeling is explained based on Figure~\ref{fig:back_lsm}.

\subsection{Vacation Queuing Model for LSM trees}
\label{sec:model}
In LSM trees, write stalls correspond to periods during which write requests are temporarily blocked. In queuing theory, such interruptions are referred to as \textit{vacations}~\cite{doshi1986queueing, levy1975utilization}. If a stall caused by flush or compaction is interpreted as the system ``taking a break,'' then the LSM write path can be naturally modeled as a vacation-based queuing system~\cite{fuhrmann1985stochastic}. Since a vacation directly corresponds to a write stall, hereafter we will refer to a stall as a vacation.

\noindent\textbf{Three Types of Vacations.}
Write stalls are categorized into three types, depending on the layer where the bottleneck occurs in the LSM~\cite{writestall,yu2023adoc, balmau2019silk}.
(1) \textit{Memtable Stall}, (2) \textit{L0 Stall}, and (3) \textit{Pending Stall}.
These three types of stall are characterized by their \textit{frequency} (how often they occur) and \textit{duration} (how long each stall lasts), both of which depend on specific system parameters such as Memtable size, L0 thresholds, and the number of background threads.

\noindent\textbf{Interdependent Pipelines.}
A key insight is that the three stages--write buffering, flushing, compaction--are not independent. Optimizations at one stage directly impact the others: \emph{(i)} increasing \textit{flush throughput} immediately reduces Memtable stalls but also injects data into L0 faster, thereby increasing L0-stall frequency; \emph{(ii)} increasing L0 compaction throughput lowers L0 stalls but pushes data downstream faster, which increases the pressure on lower levels (raising pending stalls); moreover, under shared CPU/IO resources, allocating more to L0 compaction can reduce effective flush throughput, thereby increasing Memtable stalls. This interdependency is precisely why a unified vacation model is necessary: it lets us reason about stalls as coupled phenomena rather than isolated events. 

\noindent\textbf{Poisson Arrival Process.}
To characterize the write arrival pattern, we observe that in production key-value workloads such as Facebook's RocksDB, the coefficient of variation (CV) of write inter-arrival times is close to 1~\cite{cao2020characterizing,matsunobu2020myrocks,dong2021rocksdb}; a CV near 1 indicates an exponential, memoryless distribution~\cite{parzen1999stochastic}. Because LSM stores receive writes from many independent and sparse client threads, the Palm-Khintchine theorem~\cite{10.5555/1096491} implies that their superposition converges to a Poisson process~\cite{wolff1982poisson}. We therefore model the LSM arrival process as Poisson with mean rate $\lambda$.

\noindent\textbf{Memtable Stall.} 
Let the write arrival rate be $\lambda$(MB/s), the size of the active Memtable be $M$(MB), and the flush processing rate be $\mu_{\text{flush}}$(MB/s). A Memtable stall occurs when the next Memtable becomes full before the flush of the previous one has completed, which happens under the condition $\lambda > \mu_{\text{flush}}$.
Accordingly, the average stall duration $E[V^{(\mathrm{Mem})}]$(sec) and the frequency $\lambda_{v}^{(\mathrm{Mem})}$(stalls/s) are given as follows:

\vspace{-8pt}
{\small
\begin{align}
E[V^{(\mathrm{Mem})}]=\frac{M}{\mu_{\text{flush}}}, \;\lambda_{v}^{(\mathrm{Mem})}=\frac{\lambda-\mu_{\text{flush}}}{M}
\end{align}
}
\vspace{-8pt}

\noindent\textbf{L0 Stall.}
In an LSM tree, L0-to-L1 compaction is triggered when the number of SSTables in L0 exceeds the compaction threshold $N_{\text{trigger}}$ (default\footnote{Hereafter, this default refers to the default configuration in RocksDB.}: 4 SSTables). 
However, if the flush rate $\mu_{\text{flush}}$ continues to exceed the L0-to-L1 compaction rate $\mu_{\text{cmp0}}(c_{0})$(MB/s) during compaction, the number of SSTables in L0 eventually reaches the write stall threshold $N_{\text{stall}}$ (default: 36 SSTables).
Here, $c_0$ denotes the number of compaction threads assigned to L0 compaction. Since L0-to-L1 compaction is typically not parallelized, $c_0 = 1$ by default.

Let $\Delta N = N_{\text{stall}} - N_{\text{trigger}}$ denote the gap between the two thresholds. To trigger an L0 stall, this gap must be filled with additional files at a net growth rate of $\mu_{\text{flush}} - \mu_{\text{cmp0}}(c_{0})$.
Assuming the average size of an SSTable is $S_{\text{SST}}$, the time required to accumulate $\Delta N \cdot S_{\text{SST}}$ corresponds to the average time between L0 stalls.
The time to drain this accumulated data at rate $\mu_{\text{cmp0}}(c_{0})$ corresponds to the stall duration.
Accordingly, the average duration $E[V^{(\mathrm{L0})}]$(sec) and frequency $\lambda_{v}^{(\mathrm{L0})}$(stalls/s) of L0 stalls are given as follows:

\vspace{-8pt}
{\small
\begin{align}
E[V^{(\mathrm{L0})}]
= \frac{\Delta N\ \cdot S_{SST}}
{\mu_{\text{cmp0}}(c_{0})},\;
    \lambda_{v}^{(\mathrm{L0})}
= \frac{\mu_{\text{flush}}-\mu_{\text{cmp0}}(c_{0})}
{\Delta N\ \cdot S_{\text{SST}}}
\;
\end{align}
}
\vspace{-8pt}

\noindent\textbf{Pending Stall.}
Let $k \cdot \mu_{\text{cmp0}}(c_{0})$ denote the rate at which data is pushed from L0 to deeper levels (Ld), and let $\mu_{\text{cmpd}}(c_{d})$(MB/s) denote the compaction rate at Ld.
Here, $k$ represents the fraction of data retained after duplicate removal during L0-to-L1 compaction.
The difference in these rates,
$k \cdot \mu_{\text{cmp0}}(c_{0}) - \mu_{\text{cmpd}}(c_{d})$,
determines the rate at which the pending compaction backlog grows.
When the backlog size exceeds a threshold $B_{\text{th}}$, a pending stall occurs.
Assuming that a pending stall is cleared once $S_{\text{pend}}$ bytes from the accumulated backlog are processed at the Ld compaction rate,
the average stall duration $E[V^{(\mathrm{Pend})}]$(sec) and frequency $\lambda_{v}^{(\mathrm{Pend})}$(stalls/s) are given as follows:

\vspace{-8pt}
{\small
\begin{align}E[V^{(\mathrm{Pend})}]=\frac{S_{\text{pend}}}{\mu_{\text{cmpd}}(c_{d})}, \;\lambda_{v}^{(\mathrm{Pend})}=\frac{k \cdot \mu_{\text{cmp0}}(c_{0})-\mu_{\text{cmpd}}(c_{d})}{B_{\text{th}}}
\end{align}
}
\vspace{-8pt}

\noindent\textbf{Base Stall Model.}
So far, we have modeled the durations of the three types of stalls using a vacation-based approach.
By incorporating these stalls into an M/G/1 queue--Poisson arrivals (M), general service time (G), and a single server (1)--the average base stall time $E[B]_{\text{base}}$ is given as follows:

\vspace{-4pt}
{\small
\begin{align}
E[B]_{\text{base}}
  &= \frac{\lambda}{1-\rho}
     \left(
     \begin{alignedat}{2} 
       &\lambda_{v}^{(\mathrm{Mem})}\,E[V^{(\mathrm{Mem})}]  
       &\;&(\text{Memtable Stall}) \\[4pt]
       &+ \lambda_{v}^{(\mathrm{L0})}\,E[V^{(\mathrm{L0})}] 
       &\;&(\text{L0 Stall}) \\[4pt]
       &+ \lambda_{v}^{(\mathrm{Pend})}\,E[V^{(\mathrm{Pend})}] 
       &\;&(\text{Pending Stall})
     \end{alignedat}
     \right) \notag \\[6pt]
\rho &= \lambda\,E[S] < 1 \label{eq:base}
\end{align}
}
\vspace{-12pt}

Here, $\lambda$ denotes the average arrival rate, and $E[S]$ represents the average service time required to safely record a request in the Memtable. The system utilization is given by $\rho = \lambda E[S]$. The term $\tfrac{\lambda}{1 - \rho}$ is a correction factor derived from the mean residual time law of the M/G/1 queue, reflecting the fact that when the server is utilized at rate $\rho$, the queuing delay is amplified by a factor of $\tfrac{1}{1 - \rho}$~\cite{cohen2012single}.

Finally, we define $\alpha$ as the probability that the LSM tree encounters a stall at an arbitrary time, and $\bar{V}$ as the mean stall duration. Based on these definitions, $\alpha$, $\bar{V}$, and $E[B]_{base}$ are expressed as follows.

\vspace{-16pt}
{\small
\begin{align}
    \alpha = \begin{bmatrix} \lambda_{v}^{(\mathrm{Mem})} \;\; \lambda_{v}^{(\mathrm{L0})} \;\; \lambda_{v}^{(\mathrm{Pend})} \end{bmatrix}
    ,\;
    \bar{V}= \begin{bmatrix}{E}[V^{(\mathrm{Mem})}] \\ {E}[V^{(\mathrm{L0})}] \\ {E}[V^{(\mathrm{Pend})}]\end{bmatrix}
\end{align}
}
\vspace{-8pt}
{\small
\begin{align}
   E[B]_{base}= \frac{\lambda}{1-\rho}\,\alpha\,\bar{V}
   \label{eq:model_final}
\end{align}
}

\noindent\textbf{Coupling and Trade-offs.}
The model makes the interdependencies explicit. From the stall-frequency expressions,

\vspace{-8pt}
{\small
\begin{align}
\partial \lambda_{v}^{(\mathrm{Mem})} / \partial \mu_{\text{flush}} = -1/M < 0, \\
\partial \lambda_{v}^{(\mathrm{L0})} / \partial \mu_{\text{flush}} = 1/(\Delta N S_{\text{SST}}) > 0
\end{align}
}
\vspace{-8pt}

Thus, (i) raising $\mu_{\text{flush}}$ reduces Memtable stalls but increases L0 stalls; and (ii) raising $\mu_{\text{cmp0}}$ reduces L0 stalls but increases pending stalls (and, under shared resource constraints, may also reduce $\mu_{\text{flush}}$, increasing Memtable stalls). In practice, $\mu_{\text{flush}}$, $\mu_{\text{cmp0}}$, and $\mu_{\text{cmpd}}$ are coupled through a bounded CPU/IO budget; allocating more threads or bandwidth to one stage typically reduces the effective throughput of another.
\vspace{-4pt}
\subsection{The Limitations of Monolithic LSM-trees}
\label{sec:motivation} 
\vspace{-4pt}

Armed with this model, we now analyze how two orthogonal state-of-the-art techniques --- Sub-Compaction~\cite{subcompaction} and ADOC~\cite{yu2023adoc} --- impact $E[B]_{\text{base}}$, yet cannot eliminate stalls. 

\noindent\textbf{Sub-Compaction.}
RocksDB reduces L0 stalls by parallelizing L0-to-L1 compaction. Each SSTable's key range is partitioned into anchor points, with threads compacting independently~\cite{subcompaction}. This mitigates the sequential bottleneck from overlapping ranges. 

Under our model, the effective throughput is: $\mu_{\text{cmp0}}^{\textsc{sc}}(c_{0}) = \frac{\mu_{\text{cmp0}}(1)}{(1 - p) + \frac{p}{c_{0}}}$, where $p$ is the parallelizable fraction. However, by Amdahl's law, throughput is capped at $
\frac{\mu_{\text{cmp0}}(1)}{1 - p}$. While Sub-Compaction lowers L0 stall frequency and duration, it also increases downstream injection rate $k\,\mu_{\text{cmp0}}$, thereby \emph{raising pending-stall frequency} $\lambda_{v}^{(\mathrm{Pend})}$ unless deeper-level compaction scales proportionally. Moreover, if compaction threads contend for shared IO/CPU, the effective $\mu_{\text{flush}}$ may drop, \emph{raising Memtable-stall frequency}. Hence, Sub-Compaction improves one stage at the cost of others. 

\noindent\textbf{ADOC.}
ADOC~\cite{yu2023adoc} adaptively tunes Memtable size $M$ and compaction threads $c=c_0+c_d$:
\vspace{-4pt}
\squishlist
\item{
Increasing $M$ reduces Memtable stall by lowering flush frequency, but yields larger $S_{\text{SST}}$, damping $\mu_{\text{cmp0}}$ and \emph{increasing} L0 stall.
}
\item {
Increasing $c_0$ accelerates L0 compaction, reducing L0 stalls but \emph{increasing} pending stalls via higher $k\,\mu_{\text{cmp0}}$ and, under contention, Memtable stalls via lower effective $\mu_{\text{flush}}$.
}
\item {
increasing $c_d$ mitigates pending stalls, but under a fixed resource budget can reduce $\mu_{\text{flush}}$ or $\mu_{\text{cmp0}}$, \emph{raising} Memtable or L0 stalls.
}
\squishend
Thus, ADOC's decoupled heuristics induce a \textit{balloon effect}: mitigating one stall type often exacerbates another. It lacks a unified mechanism to eliminate stalls altogether.

\begingroup
\setlength{\textfloatsep}{0pt}  
\begin{figure}[t]
  \centering
  \captionsetup[subfigure]{labelformat=empty}
  \subfloat[]{%
    \includegraphics[width=1\linewidth]{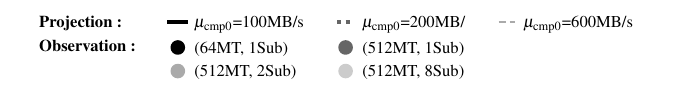}}
  \vspace{-22pt}
  \\[0pt]

  \hspace{-6pt}
  \includegraphics[width=0.45\linewidth]{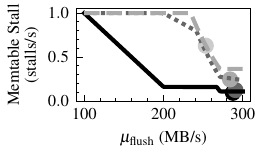}
  \hspace{0pt}
  \includegraphics[width=0.45\linewidth]{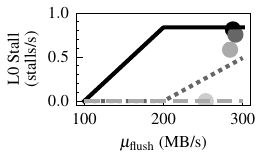}
  \par\vspace{-6pt}
  \makebox[1\linewidth]{
  \small{(a) Memtable and L0 Stall with L0 compaction throughput ($\mu_{\text{cmp0}}$)
}
  }

    \subfloat[]{
    \includegraphics[width=0.87\linewidth]{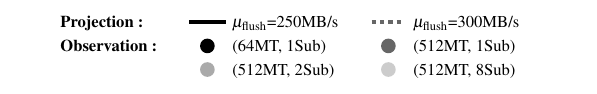}}
    \vspace{-22pt}
  \hspace{-6pt}
  \includegraphics[width=0.45\linewidth]{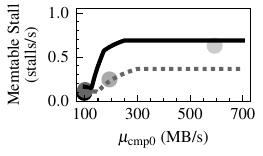}
  \hspace{0pt}
  \includegraphics[width=0.45\linewidth]{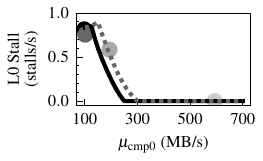}
  \par\vspace{-6pt}
  \hspace{-6pt}
  \makebox[1\linewidth]{\small{(b) 
  Memtable and L0 Stall with flush throughput ($\mu_{\text{flush}}$)}
  }

  \caption{\small Model predictions and actual observations of write stalls under different flush and compaction throughputs.
  }
  \vspace{-4pt}
  \label{fig:model_provision}
\end{figure}
\endgroup

\begin{table}[b]
    \centering 
    \caption{
    Validation of the Model.
    }
    \vspace{-2pt}
    \renewcommand{\arraystretch}{1.2}
    \vspace{-2pt}
    \resizebox{\columnwidth}{!}{
    \scriptsize{
\begin{tabular}{cc|cccc}
\toprule
\multicolumn{2}{c|}{\textbf{Stall (stalls/s)}} & \textbf{64MT, 1Sub} & \textbf{512MT, 1Sub} & \textbf{512MT, 2Sub} & \textbf{512MT, 8Sub} \\ \midrule
\multicolumn{1}{c|}{\multirow{2}{*}{\textbf{MT Stall}}} & Obs.  & 0.094 & 0.12 & 0.24 & 0.63 \\  
\multicolumn{1}{c|}{}                                   & Pred. & 0.11  & 0.11 & 0.24 & 0.63 \\  
\midrule
\multicolumn{1}{c|}{\multirow{2}{*}{\textbf{L0 Stall}}} & Obs.  & 0.81  & 0.75 & 0.58 & 0    \\  
\multicolumn{1}{c|}{}                                   & Pred. & 0.83  & 0.83 & 0.45 & 0    \\  
\bottomrule
\end{tabular}
    }
    }
    \label{tbl:model_errer_rate}
\end{table}

\noindent\textbf{Preliminary Modeling Results and Analysis.}
The proposed write stall model enables us to understand how the interaction between flush and compaction produces stalls.
Figure~\ref{fig:model_provision} presents the model’s predictions and the results obtained from RocksDB using the db\_bench's FillRandom 8 KB workload(same as \S~\ref{sec:eval}) under different flush and compaction throughputs.
The lines indicate the predicted values of our write stall model, while the circles represent actual measurements under varying system parameters.

To clearly observe the interaction between the two dominant types of write stalls---Memtable stalls and L0 stalls---we conducted experiments under conditions where pending stalls do not occur.
In all graphs, the $y$-axis reports the average ratio of stall duration to total execution time, expressed as stalls per second (stalls/s).

Figure~\ref{fig:model_provision}(a) shows stall trends when $\mu_{\text{cmp0}}$ is fixed and $\mu_{\text{flush}}$ increases.
Here, increasing $\mu_{\text{flush}}$ along the $x$-axis corresponds to changes in Memtable size under ADOC.
As the three observed points suggest, lighter circles indicate reduced Memtable stalls but increased L0 stalls.
In other words, while the degree of Memtable and L0 stalls varies with system parameters, the overall stalls never disappear—a balloon effect is observed.
The measurements lie close to the model’s predictions for the three values of $\mu_{\text{cmp0}}$.

Figure~\ref{fig:model_provision}(b) depicts stall behavior when $\mu_{\text{flush}}$ is fixed and $\mu_{\text{cmp0}}$ increases, as the number of Sub-Compaction threads grows.
Here, the trend reverses: L0 stalls decrease while Memtable stalls increase, revealing the balloon effect.  All measurements closely match the model’s predictions for the two values of $\mu_{\text{flush}}$. 
\textit{This balloon effect reveals a fundamental limitation: optimizations such as ADOC or Sub-Compaction only shift the stall bottleneck between levels, but can never eliminate stalls entirely.} 

\noindent\textbf{Model Validation.}
The accuracy of the proposed model is validated in the contexts of Sub-Compaction and ADOC. Validation is based on the experimental and observed values presented in Figure~\ref{fig:model_provision}. 
Table~\ref{tbl:model_errer_rate} compares predicted and observed write stall rates. The model achieves a mean absolute error (MAE) of 0.03~stalls/s in predicting both Memtable and L0 stall rates, while consistently capturing the balloon effect between the two stall types.

\vspace{-7pt}
\subsection{Opportunity: LSM Sharding}
\label{sec:opportunity}
\vspace{-3pt}

The preceding analysis shows that in a monolithic LSM pipeline, stalls cannot be eliminated regardless of parameter tuning, since the arrival rate $\lambda$ eventually exceeds the internal throughput ($\mu_{\text{flush}}$, $\mu_{\text{cmp0}}$, $\mu_{\text{cmpd}}$). A structural alternative is to partition the LSM tree into $L$ independent shards, reducing the per-shard arrival rate to $\lambda/L$. This slows Memtable filling, staggers flushes, and lowers utilization to $\rho/L$. 

If a shard stalls with probability $\alpha$, then the probability of all $L$ shards stalling simultaneously is $\alpha^L$, which decreases exponentially. 
The expected stall time after sharding is thus approximated by:

\vspace{-10pt}
{\small
\begin{align}
    E[B]_{L}\;\simeq\;\frac{\lambda}{1-\rho/L}\,\alpha^{L}\,\bar{V}.
\end{align}
}
\vspace{-10pt}

Unlike Sub-Compaction or ADOC, which only shift stall types, sharding reduces stall probability exponentially and provides a structural path toward near-stall-free operation.

\begin{figure}[H]
\vspace{-10pt}
	\centering
    \includegraphics[width=0.55\linewidth]{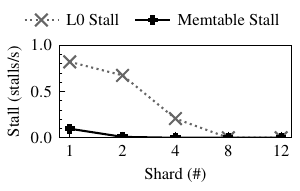}
    \vspace{-6pt}
    \caption{\small Changes in stall according to the number of shards based on the model in the preliminary results (512MT, 1 Sub-Compaction, 8KB FillRandom). 
	}
    \vspace{-8pt}
	\label{plot:model_sharding}
\end{figure}

Figure~\ref{plot:model_sharding} shows model-based theoretical changes in stalls as a function of the number of shards under preliminary results (512MT, 1 Sub-Compaction, 8KB FillRandom). As the number of shards increases, stalls decrease sharply and approach nearly zero at eight shards. This behavior is validated by the evaluation results ($\S$\ref{eval:perf_write}).

\vspace{-10pt}
\begin{figure}[H]
	\centering
    \includegraphics[width=0.7\linewidth]{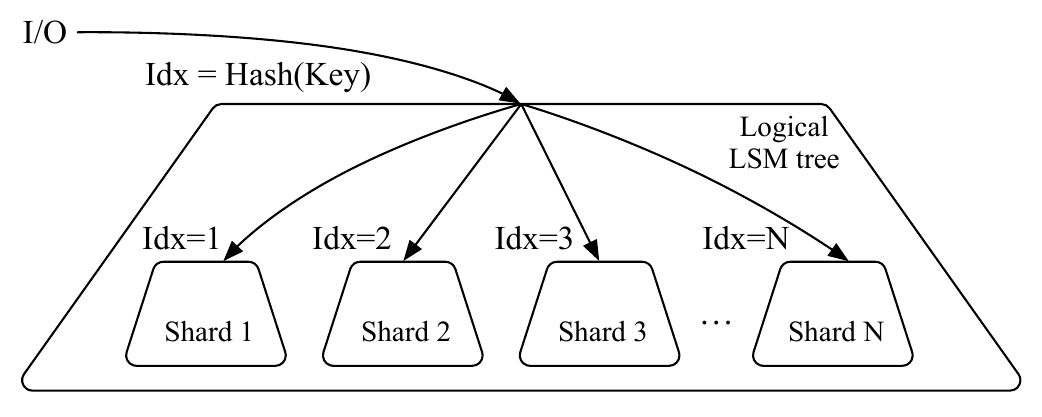}
    \vspace{-6pt}
    \caption{\small  A high-level concept of \tbdfixed{}.
    }
   \vspace{-4pt}
	\label{fig:sharding}
\end{figure}

\section{Design of \tbdfixed{}}
\label{sec:design}
\vspace{-2pt}

\tbdfixed{} adopts a shard-based LSM architecture that departs from the traditional monolithic design by assembling multiple smaller and independent LSM shards. Each shard manages part of the global key space and executes its own compaction processes in isolation, effectively distributing write traffic and alleviating compaction-induced stalls. At a higher level, the full LSM tree can be regarded as a logical aggregation of these shards, as shown in Figure~\ref{fig:sharding}. To support this architecture, \tbdfixed{} incorporates directory services to track the placement of key-value pairs across shards, ensures global key ordering, and maintains the same level of failure consistency as a conventional non-sharded LSM tree.

\noindent\textbf{Model-based Intuition for Number of Shards.}
The modeling results provide important intuition on how the number of shards affects write stall mitigation (\S\ref{sec:opportunity}). Based on observable parameters in the LSM pipeline, including the write ($\lambda$), flush ($\mu_{\text{flush}}$), and compaction rate ($\mu_{\text{cmp0}}$, $\mu_{\text{cmpd}}$), the model quantitatively estimates how stall probability decreases as the number of shards increases. 
This helps approximate the necessary range of shards required to effectively mitigate write stalls in a given environment, but it cannot guarantee identical results when applied to an actual system.

Accordingly, the design of \tbdfixed{} does not attempt to automatically tune or determine the optimal number of shards; instead, it focuses on experimentally demonstrating that \tbdfixed{} can practically mitigate write stalls across diverse shard configurations and system conditions (\S\ref{sec:eval}). A more detailed discussion of shard-count trade-offs is deferred to the Discussion (\S\ref{sec:discuss}).

\vspace{-8pt}
\subsection{Design Goals}
\vspace{-2pt}
\squishlist
\item 
\textbf{Objective \#1: Practical Reduction of Write Stalls.} 
\tbdfixed{} introduces a \textit{primary LSM shard} and one or more \textit{fallback LSM shards} for each key-value pair, effectively mitigating write stalls by bypassing writes even when the internal throughput of LSM shards is insufficient.

\item
\textbf{Objective \#2: Maintain Read I/O Performance.} 
Because \tbdfixed{} consists of several small physical LSM trees, redirected KV pairs must be efficiently located. 
This requires an indexing mechanism—similar to a lightweight directory service—whose lookup cost remains minimal. 

\item
\textbf{Objective \#3: Support Global Ordering of KV Pairs.}
Keys distributed across different shards lack a global order. 
Since range queries rely on ordering, \tbdfixed{} preserves global order semantics as in a monolithic LSM tree. 

\item
\textbf{Objective \#4: Provide Consistency.} 
When KV pairs are written to fallback shards during stalls, crashes may occur anytime.
It is critical to reliably track such redirections, ensuring recovery without data loss or inconsistency. 
\squishend

\begingroup
\setlength{\textfloatsep}{-10pt} 
\begin{figure*}[t]
	\centering
    \includegraphics[width=0.9\linewidth]{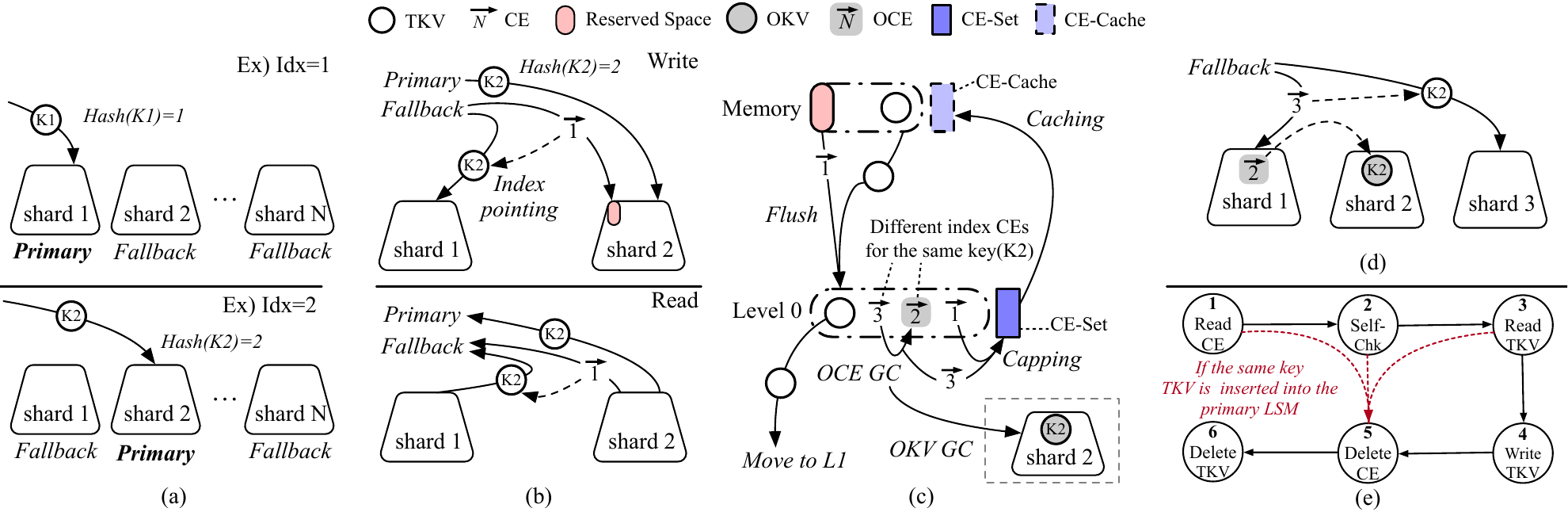}
    \vspace{-4pt}
    \caption{\small Architecture and operations of \tbdfixed{}. (a) Examples of the primary LSM and fallback LSM for each key, (b) write and read paths, (c) internal operations of LSM shard: CE Capping, CE Caching, CE/TKV Garbage Collection, (d) A scenario where obsolete data occurs, (e) Flow of performing a rollback.
    }
   \vspace{-10pt}
	\label{fig:design_overview}
\end{figure*}
\endgroup

\vspace{-10pt}
\subsection{Dynamic Rerouting with Primary-Fallback}
\vspace{-2pt}

Figure~\ref{fig:design_overview}(a) illustrates how primary and fallback shards are determined. 
\tbdfixed{} adopts hash-based placement: the primary shard index for a key is obtained by hash(key), mod(num\_shards). 
Each shard maintains its own Memtable, flushes it when full, and performs leveled compactions (including L0-to-L1) in the same manner as a monolithic LSM tree.  
For example, if $\langle K1 \rangle$ maps to shard~1, then shard~1 serves as the primary shard, while all other shards act as fallbacks, as shown in Figure~\ref{fig:design_overview}(a). 
This mapping is per-key: a shard may be primary for some keys and fallback for others. 

\subsubsection{Write Procedure}
Figure~\ref{fig:design_overview}(b) shows the write path for the $\langle K2 \rangle$ example, divided into  \textit{Primary} and  \textit{Fallback}:

\vspace{-2pt}
\begin{enumerate}[leftmargin=12pt,label=\textbf{\arabic*)},itemsep=-2pt,topsep=2pt]
\item
\textbf{Primary path.} 
If the primary shard of $\langle K2 \rangle$ is not stalled, the write is inserted directly into it. 

\item
\textbf{Fallback path.} 
If the primary shard is stalled, the write redirected to a non-stalled fallback shard (e.g., shard~1).

\item
\textbf{CE and TKV.} 
In the fallback case, an indexing entry is inserted in the primary shard to record the redirection. This indexing entry is referred as \textit{Clue-Entry} (CE), while the original KV pair for $\langle K2 \rangle$ is called the \textit{True-KV}. The CE entry in primary shard points to the fallback shard containing the TKV, e.g., $\langle K2, \overrightarrow{1} \rangle$. This approach leads to versioning and obsolescence issues which covered in detail in Section~\ref{sec:design_internal}.

\item
\textbf{Reserved Space for CE.} 
Since the primary shard is stalled, regular TKV entries cannot be inserted. However, CEs are only a few bytes. To guarantee CE insertions, \tbdfixed{} reserves 2~MB of Memtable space in each shard, excluded from stall threshold calculations. Thus, CEs can always be inserted even during stalls. With 20-byte entries (16-byte key, 4-byte pointer), this space holds up to 100,000 CEs. Note that the reserved space size is user-defined.
\end{enumerate}

\noindent\textbf{Fallback Shard Selection Policy.}
Since fallback shards can also experience stalls, their selection must be handled carefully. In a system with $N$ shards, candidates are chosen from the $N-1$ non-stalled shards. Selection follows a strict priority order: (1) smallest L0 size; (2) lowest Memtable usage; (3) smallest pending backlog; and (4) smallest total data size. This approach balances load across shards and reduces the risk of cascading stalls.
Redirecting I/O to a fallback shard and inserting the corresponding CE incur extra reads and writes, which can increase read, write, and space amplification. These impacts are evaluated in detail in Section~\ref{sec:eval_feature}.

\subsubsection{Read Procedure}
The read path in Figure~\ref{fig:design_overview}(b) is also divided into two cases: 
\vspace{-2pt}
\begin{enumerate}[leftmargin=12
pt,label=\textbf{\arabic*)},itemsep=-2pt,topsep=2pt]
\item
\textbf{Primary path.} 
If the primary shard holds the TKV, it is returned immediately.

\item
\textbf{Fallback path.} 
If the  primary shard returns a CE, the pointer is followed to the fallback shard to retrieve TKV. This requires two reads across shards, doubling worst-case latency. To mitigate this, \tbdfixed{} employs \textit{CE capping} and \textit{CE caching} in Section~\ref{sec:design_internal}.
\end{enumerate}

\subsection{Managing Clue Entries}
\label{sec:design_internal}
Figure~\ref{fig:design_overview}(c) shows an LSM shard's internal structure and operation. Each shard behaves like a standalone LSM tree, following traditional LSM principles. To enable efficient interaction between TKVs and CEs, \tbdfixed{} introduces additional components: a reserved space and a CE-cache in memory, and a CE-Set at L0. These structures ensure that CEs are tracked and accessed efficiently during both writes and reads. 

When a flush occurs, the CE from the reserved space and the TKV from the Memtable are written together into an L0 SSTable. During L0 compaction, they are split, TKVs are compacted into L1, while CEs are retained in L0 via a process called \textit{CE L0 Capping}, which ensures CEs never move beyond L0. To further reduce lookup latency, the CE-Set is mirrored in memory via \textit{CE Caching}. Both the CE-Set and CE-cache are updated after every L0 compaction. The system performs lookups in the following order: (1) Memtable, (2) CE-cache, (3) L0 SSTables, and (4) CE-Set. 

\noindent\textbf{Garbage Collection.}
When multiple CEs exist for the same key, older versions become obsolete. For example, in Figure~\ref{fig:design_overview}(d), if $\langle K2, \overrightarrow{2} \rangle$ is inserted and later replaced by $\langle K2, \overrightarrow{3} \rangle$, the former becomes an obsolete CE (OCE). The KV pair it points to --- stored in another LSM shard --- becomes an obsolete KV (OKV) pair. Without cleanup, these outdated entries would remain in the system indefinitely, consuming unnecessary storage. To address this, garbage collection (GC) is build into the L0 compaction process (Figure~\ref{fig:design_overview}(c)), where both OCE and OKV pair cleanup are performed. 

OCE garbage collection occurs naturally during L0 compaction, which keeps only the most recent CE per key. For OKV pairs, since they are stored in different shards, \tbdfixed{} sends a delete request to the relevant shard whenever its corresponding OCE is removed. This ensures obsolete KV pairs are properly cleaned up. Because GC is integrated into existing compaction, its overhead is typically minimal. Only in rare workloads, with extremely frequent updates to the same key might the GC load become significant enough to impact performance -- scenarios that are uncommon in practice. 

\subsection{Managing Rollbacks}
\label{sec:rollback}
\tbdfixed{} temporarily redirects KV pairs to fallback shards when a primary shard enters a write stall. To maintain consistency across the LSM tree, these redirected KV pairs must eventually be written back to their original shard through a rollback process. Rollbacks are performed by dedicated background threads, triggered independently in each shard based on specific policies. 

Figure~\ref{fig:design_overview}(e) shows the rollback process as follows. Step~\circled{1} Read a CE (e.g., $\langle K2, \overrightarrow{2} \rangle$) from the CE-cache of the primary shard. Step~\circled{2} Perform a self-check to see if the corresponding TKV has already been inserted into primary shard. Step~\circled{3} If not found, retrieve the TKV from the fallback shard (e.g., shard~2). Step~\circled{4} Write the TKV back to the primary shard. Step~\circled{5} -- \circled{6} Delete the CE from the primary shard and the TKV from the fallback shard. However, this rollback flow assumes a static system state. In practice, new writes may occur concurrently with rollback, introducing the risk of inconsistencies --- most notably, a version inversion problem. 

\noindent\textbf{Handling Version Inversion Issues.}
\label{sec:version_inversion}
During rollback, foreground I/O thread may continue working to the primary LSM shard. This can cause a version inversion issue, where a newly written TKV is overwritten by an outdated version from rollback. For instance if a new TKV for key $K2$ is inserted into the primary shard between Step~\circled{1} and Step~\circled{3}, and the rollback process proceeds without recognizing this update, it may overwrite the new TKV. To prevent this, \tbdfixed{} checks for key conflicts. If a new TKV for the same key is detected, the rollback skips re-insertion and proceeds directly to cleanup (Steps~\circled{5} -- \circled{6}), ensuring data correctness. 

\noindent\textbf{Policy of Rollback Triggering.}
Rollback introduces additional write/read load to both primary and fallback shards. To avoid degrading performance, rollback is only triggered under controlled conditions. \tbdfixed{} uses two trigger policies. The \textit{stabilization-state} trigger runs rollback only when all shards are stable --- i.e., no active compactions and a small L0 --- indicating low write activity. If a new write burst disrupts this state, rollback is aborted. The \textit{CE threshold} trigger activates rollback when the number of CEs exceeds a predefined limit (default: 1 million). This prevents excessive CE build up from impacting reads. When this trigger fires, rollback proceeds regardless of shard state, and no further CEs are accepted, even at the cost of potential write stall.

\subsection{Range Query Processing in \tbdfixed{}}
\label{sec:design_range_query}
In a shard-based LSM architecture, there is no global ordering across shards. A naive approach to range queries would involve comparing the current key of every shard during each call to \texttt{Next()}. However, this method introduces significant overhead, as it requires serial comparisons across all shards. As the number of shards increases, this overhead grows linearly, degrading range query performance to below that of a monolithic LSM tree. 

To address this, DIALSM implements a parallel range query mechanism that considers the start key 
($K$) and the scan length ($L$) as scan bounds and exploits the parallelism of the sharded structure. 
Instead of scanning serially, all shard perform \texttt{Seek($K$)} concurrently. Each shard then independently invokes \texttt{Next()} until it reaches the end of its local portion of the query range ($K+L$). Because data is hash-partitioned across shards, each shard holds only a subset of keys. As the number of shards increases, the key span handled by each shard shrinks, resulting in improved range query throughput. 

Algorithm~\ref{algo:parallelrange} outlines this process. Given a user-specified start key $K_{start}$ and scan length $L$, the system first computes the upper bound key $K_{end}$ (lines 1--3). Then, each shard concurrently executes \texttt{Seek($K_{start}$)}. If the key is not present in a shard, its iterator advances to the nearest key greater than or equal to $K_{start}$ (line 4--5). From there, each shard continues scanning by repeatedly invoking \texttt{Next()} until it reaches $K_{end}$ (line 6--8). This entire process runs in parallel, with each shard operating independently. Once all shards complete local scans, the range query is complete (line 9).

While parallel range queries may increase CPU usage due to concurrent shard execution, this is offset by a significant reduction in overall processing time. As a result, \tbdfixed{} achieves efficient and scalable range query performance.

\begingroup
\setlength{\textfloatsep}{-10pt}   
\begin{algorithm}[t]
\scriptsize
\DontPrintSemicolon
\SetKwFunction{Frangescan}{RangeScan}
\SetKwProg{Fn}{Function}{:}{}

\Fn{\Frangescan{$K_{start}, L$}}{
    $K_{end} \gets K_{start} + L$ \; 
    $Output \gets [\;]$ \;
    \ForEach{$s \in Shards$ in parallel}{ 
        $Shard\_iter_s.Seek(K_{start})$ \; 
        \While{$Shard\_iter_s.Valid()$ {\bf and} $Shard\_iter_s.GetKey() < K_{end}$}{
            $Output \gets Shard\_iter_s.GetKey(), Shard\_iter_s.GetValue()$ \;
            $Shard\_iter_s.Next()$ \;
        }
    }
    \KwRet{$Output$}
}
\caption{Parallel Range Scan in \tbdfixed{}}
\label{algo:parallelrange}
\end{algorithm}
\endgroup

\vspace{-4pt}
\subsection{Handling Failures and Recovery}
\label{sec:consistency}

System crashes can lead to consistency issues in \tbdfixed{}. A full cross-validation of all LSM shards would ensure correctness, but this approach is prohibitively expensive due to the large number of reads required. Instead, \tbdfixed{} employs a fast recovery method that avoid global cross-validation and enables efficient recover after a crash. In the recovery phase, user I/O is not accepted.

\noindent\textbf{Failures in Writing.}
During normal operation, a failure may occur while inserting a KV pair into the fallback LSM shard, typically triggered by a write stall in the primary shard. We define this as Failure Case-1 (FC-1): 
\squishlist
\item \textbf{FC-1:} 
The TKV is successfully inserted into the fallback shard, but a crash occurs before the corresponding CE is written to the primary shard. 
\squishend 

In FC-1, the fallback shard contains a TKV that is effectively invisible due to the missing CE entry in primary shard. To recover, \tbdfixed{} reads the most recent TKV in each shard, hashes the key to determine its rightful primary shard, and checks for a matching CE entry. If no CE is found in the primary shard, the system assumes an incomplete write and rolls the TKV back to the primary shard --- without creating a CE --- thereby restoring consistency. 

At this point, consecutive TKVs with missing CEs may have been inserted into the fallback shard. Therefore, this process is repeated until no TKVs with missing CEs remain.

\noindent\textbf{Failures in the Process of Rollback.}
Failures may occur during the rollback process. We consider two such cases:
\squishlist
\item \textbf{FC-2:} A crash occurs after inserting the TKV into the primary shard (Step~\circled{4}) but before removing the CE from the CE-Set (Step~\circled{5}).
\item \textbf{FC-3:} A crash occurs after removing the CE (Step~\circled{5}) but before deleting the TKV from the fallback shard (Step~\circled{6}).
\squishend

In FC-2, both TKV and CE exist in the system, which can cause unnecessary reads and increased latency. To resolve this, \tbdfixed{} checks if a shard contains both a recently inserted TKV and a matching CE. If so, the rollback is resumed and completed that point. For FC-3, the same KV pair exists in both primary and fallback shards, but the CE has already been deleted --- making it unclear which shard the fallback was. In recovery, each shard scans for its most recent TKVs and checks all other shards for duplicates. The shard that still holds the redundant KV (identified via key hashing) is treated as the fallback, and its KV is removed to finalize rollback. 
\vspace{-4pt}
\section{Evaluation}
\label{sec:eval}

\subsection{Experimental Setup}
\label{sec:eval_setup}

\noindent\textbf{Implementation.} 
We implemented \tbdfixed{} on top of RocksDB v8.3.2, a widely adopted LSM tree based key-value store. To enable this, we introduced the db\_master class, which ensures that each LSM shard in \tbdfixed{} operates as an independent RocksDB instance. The db\_master class exposes APIs for operations, including Open(), Close(), Put(), Get(), and Seek(), and supports the following functionality. To avoid write stalls and select the optimal fallback LSM shard, the system periodically monitors shard states, including Memtable usage, L0 size, and the compaction backlog. In our experiments, the monitoring interval is set to 0.1 seconds.

 The source code for \tbdfixed{} is publicly available at \url{https://github.com/DISCOS-LAB/DiaLSM}.
 
\noindent\textbf{Platform.}
Our experiments were conducted on a server equipped with two AMD EPYC 7352 24-Core CPUs, 256~GB DDR4 DRAM, and a PCIe Gen3x4 3.84~TB Samsung MZSLR3T8HBLS-00A07 NVMe SSD.
The system ran CentOS 7 with the ext4 file system.

\begin{table}[!b]
    \centering
     \renewcommand{\arraystretch}{1.1}
    \caption{\small Configuration details of evaluation candidates. 
    }
    \resizebox{0.45\textwidth}{!}{
\begin{tabular}{c|ccc}
\toprule
 & \multicolumn{1}{c}{\textbf{\tbdfixed{}($n$)}} & \multicolumn{1}{c}{\textbf{ADOC($m$)}} & \textbf{RocksDB($m$)} \\ \midrule
\textbf{Each Memtable Size (MB)} & \multicolumn{1}{c}{$512 / n$} & \multicolumn{1}{c}{64 - 512} & $512 / m$ \\ 
\textbf{Memtable Count (\#)} & \multicolumn{1}{c}{$2n$} & \multicolumn{1}{c}{2} & $2m$ \\ 
\textbf{Total Memtable Size (MB)} & \multicolumn{1}{c}{1024} & \multicolumn{1}{c}{1024} & 1024\\ 
\textbf{Flush Thread (\#)} & \multicolumn{1}{c}{$2n$} & \multicolumn{1}{c}{2} & $2m$ \\ 
\textbf{Compaction  Thread (\#)} & \multicolumn{1}{c}{$n \cdot \lceil 32 / n \rceil $} & \multicolumn{1}{c}{32} & 32 \\ 
\textbf{Max L0 SST (\#)} & \multicolumn{1}{c}{$36 \cdot n$} & \multicolumn{1}{c}{36} & $36 \cdot m$\\ 
\textbf{Total L0 Size (MB)} & \multicolumn{1}{c}{$512 \times 36$} & \multicolumn{1}{c}{$512 \times 36$} & $512 \times 36$\\ 
\textbf{L$i$ Size ($i\ge1$) (MB)} & \multicolumn{1}{c}{$(256 \times 10^{i}/n) \cdot n$} & \multicolumn{1}{c}{$256 \times 10^{i}$} & $256 \times 10^{i}$\\ 
\bottomrule
\end{tabular}
    }
    \label{tbl:eval_system}
\end{table}

\begingroup
\setlength{\textfloatsep}{-20pt}
\begin{figure*}[t]
  \centering

  \captionsetup[subfigure]{labelformat=empty}
  \subfloat[]{%
    \includegraphics[width=0.5\linewidth]{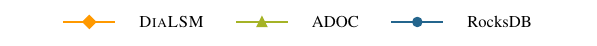}}
  \vspace{-20pt}
  \\[0pt]

  \hspace{-4pt}
  \includegraphics[width=0.165\linewidth]{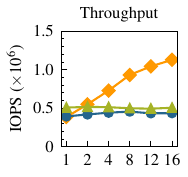}
  \hspace{-7pt}
  \includegraphics[width=0.16\linewidth]{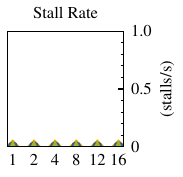}
  \hspace{-7pt}
  \includegraphics[width=0.165\linewidth]{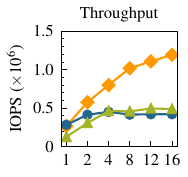}
  \hspace{-7pt}
  \includegraphics[width=0.16\linewidth]{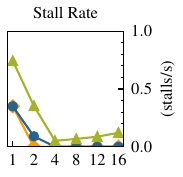}
  \hspace{-7pt}
  \includegraphics[width=0.165\linewidth]{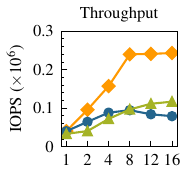}
  \hspace{-7pt}
  \includegraphics[width=0.16\linewidth]{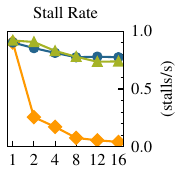}

  \vspace{-4pt}
  {\small\makebox[\linewidth][c]{\footnotesize Shard (\#) or Sub Comp. Thread (\#)}}
\vspace{-10pt}
  
  \makebox[0.33\linewidth]{\small{(a) 128B (Small)}}
   \hspace{-7pt}
  \makebox[0.33\linewidth]{\small{(b) 1KB (Medium)}}
  \hspace{-7pt}
  \makebox[0.33\linewidth]{\small{(c) 8KB (Large)}}

  \vspace{-4pt}
  \caption{\small Comparison of throughput and stall for different value sizes on the FillRandom workload. On the $x$-axis, \tbdfixed{} refers to the number of LSM shards, and all other systems refer to the number of Sub-Compaction threads.}
  \vspace{-20pt}
  \label{fig:eval_fr_system_compair}
\end{figure*}
\endgroup

\noindent\textbf{Comparison Systems.}
We compare ADOC~\cite{yu2023adoc}, RocksDB~\cite{rocksdb}, with Sub-Compaction enabled, and \tbdfixed{}.
Configurations of ADOC, RocksDB and \tbdfixed{} are as follows. 

\squishlist
\item 
\textbf{ADOC}: 
We have enabled the Sub-Compaction in ADOC under RocksDB v8.3.2. Note that ADOC dynamically expands Memtables up to 512~MB. ADOC($m$) refers to an ADOC with $m$ Sub-Compactions enabled. 

\item \textbf{\tbdfixed{}}: 
Each shard has two Memtables and two flush threads.
\tbdfixed{}($n$) refers to a \tbdfixed{} that utilizes $n$ LSM shards.

\item \textbf{RocksDB}:  
RocksDB v8.3.2 with Sub-Compaction enabled. RocksDB($m$) notation follow the same as ADOC.
\squishend

\noindent\textbf{Comparison Fairness.}
As shown in Table~\ref{tbl:eval_system}, all systems use the same resources, and fairness is as follows. 
\squishlist
\item \textbf{Memtable and Flush Parallelism}: As the number of shards increases, \tbdfixed{}($n$) increases Memtable($2n$) and Flush($2n$) parallelism; RocksDB($m$) is configured accordingly to match this behavior. In contrast, ADOC dynamically adjusts Memtable sizes, making equivalent configuration infeasible.

\item \textbf{Number and Size of L0 SSTables}: The L0 SSTable size depends on the Memtable size. As Memtable sizes decrease, \tbdfixed{} and RocksDB can accommodate more SSTables in L0. However, all systems use the same L0 capacity (RocksDB default: up to 36 SSTables in L0, totaling $36 \times 512$).

\item \textbf{Number of Compaction Threads}: All systems are configured with a total of 32 compaction threads. In \tbdfixed{}($n$), each shard is allocated $\lceil 32 / n \rceil$ compaction threads, so that the sum of compaction threads across all shards is 32.

\item \textbf{LSM Size}: With more shards, \tbdfixed{}($n$) reduces per-level($i$) sizes ($256 \times 10^{i}/n$) so that each shard becomes proportionally smaller, while the total LSM size remains the same as in the other systems. The L1 size (256MB) and level ratio (10) are RocksDB defaults.
\squishend

\noindent\textbf{Common RocksDB Parameter Configurations.}
RocksDB provides I/O rate throttling thresholds that throttle the I/O when the number of SSTs in L0 reaches the Write Stall Threshold (WST) during the high write I/O phase.  
To accurately assess performance degradation caused by write stalls, we disable the throttling threshold in all experiments. 
We also disabled commit logging to accurately identify only the cause of write stalls in IO.
RocksDB also provides a block cache, \tbdfixed{} with multiple LSM shards uses more cache than RocksDB and this can impact performance. We disable block cache for all systems for a fair evaluation. 

\noindent\textbf{Workloads.} To evaluate the efficacy of \tbdfixed{} in a variety of workloads, we consider various workloads from db\_bench~\cite{dbbench}, YCSB benchmark~\cite{ycsb}, and Sysbench~\cite{sysbench}. 
\squishlist
\item \textbf{FillRandom}: 100\% write workloads are evaluated in db\_bench. To emulate different arrival rates, three value sizes—128B, 1KB, and 8KB—are considered, representing very small, medium, and very large values based on prior analyses of real-world key-value workloads~\cite{cao2020characterizing, yu2023adoc, balmau2019silk, wang2023rbc, atikoglu2012workload, 7208288, yang2021large, sun2023improving}. For each I/O thread inserts 10 million entries.

\item \textbf{RRWR}: The ReadRandomWriteRandom workload supports execution under different read/write ratios. The database is first populated with 100 million (10M $\times$ 10) entries of 1KB values using FillRandom, after which each I/O thread performs read and write operations on an additional 10 million 1KB entries.

\item \textbf{YCSB}: The YCSB benchmark, a widely adopted workload for evaluating real-world scenarios, is used in the evaluation. The experiments are conducted with a value size of 1KB, and the details of each workload are presented in Table~\ref{tbl:workload_ycsb}.

\item \textbf{OLTP}: To evaluate effectiveness in RocksDB-based RDBMSs, the OLTP workloads of Sysbench~\cite{sysbench} are used. Detailed workload descriptions are provided in Table~\ref{tbl:workload_oltp}.
\squishend 

All db\_bench and YCSB workloads use 16-byte keys and operate with 10 I/O threads, except for the YCSB-E workload. In YCSB-E, a single I/O thread was used to avoid the parallel processing overhead of \tbdfixed{}, meaning that a single I/O thread performs range scans.

\subsection{Write-Only Performance}
\label{eval:perf_write}

Figure~\ref{fig:eval_fr_system_compair} analyzes throughput and stall behavior across different value sizes. With small values (128B), stalls are negligible in all systems, but \tbdfixed{} still benefits from sharding, showing throughput gains even without stall pressure. For medium values (1KB), stalls begin to appear; here, \tbdfixed{} maintains its advantage, while ADOC and RocksDB remain constrained by L0 bottlenecks.

With large values (8KB), stalls become dominant: ADOC and RocksDB demonstrate up to $17.6\times$ higher stall rates than \tbdfixed{} (94.3\% lower), limiting their throughput, whereas \tbdfixed{} achieves up to 2.4$\times$ higher throughput than ADOC by both parallelizing the pipeline and avoiding stalls. Notably, \tbdfixed{}’s throughput scales with the number of shards only up to 8, after which the bottleneck shifts from stalls to software overhead, indicating that the system reaches a saturation point beyond which additional shards bring no further gains.
In addition, ADOC sustains throughput improvements as more Sub-Compaction threads are added, while RocksDB suffers a throughput decline beyond 8 threads due to diminishing Sub-Compaction efficiency. This drop arises because smaller Memtables generate smaller SSTables in L0, reducing the benefit of partitioning. In contrast, \tbdfixed{} nearly eliminates stalls altogether, highlighting its structural advantage over both ADOC and RocksDB.

\begingroup
\setlength{\textfloatsep}{-10pt}
\begin{figure}[t]
  \centering
  \captionsetup[subfigure]{labelformat=empty}
  \subfloat[]{%
    \includegraphics[width=1\linewidth]{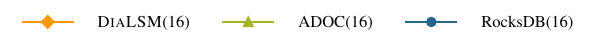}}
  \vspace{-20pt}
  \\[0pt]

  \hspace{-3pt}
  \includegraphics[width=0.325\linewidth]{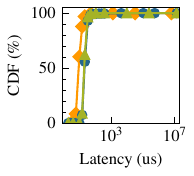}
  \hspace{-7pt}
  \includegraphics[width=0.32\linewidth]{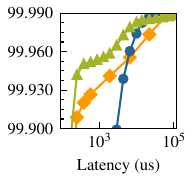}
  \hspace{-7pt}
  \includegraphics[width=0.32\linewidth]{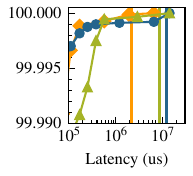}
\vspace{-4pt}
  
  \hspace{-3pt}
  \makebox[0.325\linewidth]{\small{(a) p0--p100}}
   \hspace{-7pt}
  \makebox[0.32\linewidth]{\small{(b) p99.9--p99.99}}
  \hspace{-7pt}
  \makebox[0.32\linewidth]{\small{(c) p99.99--p100}}

  \caption{\small  Cumulative distribution function (CDF) of latency for \tbdfixed{}(16), ADOC(16), and RocksDB(16) under the FillRandom workload with 8KB values. In (c), the legend’s colored lines mark latency at the p99.9999.}
  \label{fig:eval_fr_system_compair_latency}
\end{figure}
\endgroup

Figure~\ref{fig:eval_fr_system_compair_latency} illustrates the latency distribution for 8KB values. \tbdfixed{}(16) consistently lowers average and median latency by about 50\% over ADOC(16) and RocksDB(16), demonstrating the benefit of sharding in reducing stall-induced delays. At high percentiles (p99.9–p99.99), \tbdfixed{} incurs extra latency from fallback redirection and CE insertion, but this overhead remains bounded. Most importantly, at the extreme tail (p99.99+), ADOC(16) and RocksDB(16) suffer severe latency spikes due to write stalls, whereas \tbdfixed{} sustains stable performance and reduced tail-latency.

\subsection{Mixed Read/Write Workloads}
\label{eval:perf_mixed}
Earlier, we demonstrated that \tbdfixed{} outperforms existing solutions under write-only workloads, where write stalls can be most severe. However, real-world environments typically mix reads and writes, so it is crucial that \tbdfixed{} also delivers competitive performance when reads dominate. To evaluate this, we varied the read/write ratio using the RRWR workload, while keeping the experimental setup identical to the previous section.

\begingroup
\setlength{\textfloatsep}{0pt}
\begin{figure}[t]
  \centering

  \captionsetup[subfigure]{labelformat=empty}
  \subfloat[]{%
    \includegraphics[width=1\linewidth]{plots/eval_system_compair_legend.pdf}}
  \vspace{-20pt}
  \\[0pt]

  \hspace{-3pt}
  \includegraphics[width=0.365\linewidth]{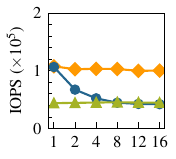}
  \hspace{-7pt}
  \includegraphics[width=0.31\linewidth]{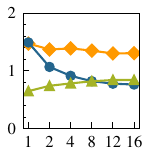}
  \hspace{-7pt}
  \includegraphics[width=0.31\linewidth]{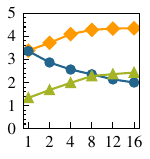}

  \vspace{-4pt}
  {\small\makebox[\linewidth][c]{\footnotesize Shard (\#) or Sub Comp. Thread (\#)}}
\vspace{-8pt}
  
  \hspace{-3pt}
  \makebox[0.365\linewidth]{\small{(a) R90/W10}}
  \hspace{-7pt}
  \makebox[0.31\linewidth]{\small{(b) R50/W50}}
  \hspace{-7pt}
  \makebox[0.31\linewidth]{\small{(c) R10/W90}}

  {\small\makebox[\linewidth][c]{{Read/Write Ratio}}}
  \caption{\small Throughput comparison for different Read/Write ratios on the ReadRandomWriteRandom workload.}
  \label{fig:eval_rrwr_system_compair}
\end{figure}
\vspace{0pt}
\endgroup

\begingroup
\setlength{\textfloatsep}{0pt}
\begin{figure}[b]
  \centering

  \captionsetup[subfigure]{labelformat=empty}
  \subfloat[]{%
    \includegraphics[width=1\linewidth]{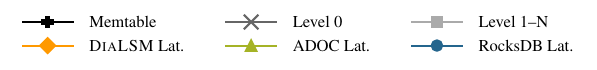}}
  \vspace{-20pt}
  \\[0pt]

  \hspace{-3pt}
  \includegraphics[width=0.345\linewidth]{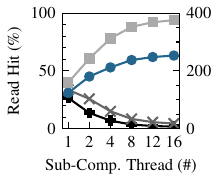}
  \hspace{-7pt}
  \includegraphics[width=0.31\linewidth]{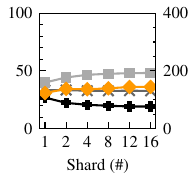}
  \hspace{-7pt}
  \includegraphics[width=0.345\linewidth]{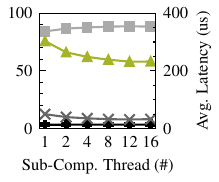}
  
  \vspace{-4pt}
  
  \hspace{-3pt}
  \makebox[0.345\linewidth]{\small{(a) RocksDB}}
  \hspace{-10pt}
  \makebox[0.31\linewidth]{\small{(b) \tbdfixed{}}}
  \hspace{-14pt}
  \makebox[0.345\linewidth]{\small{(c) ADOC}}

  \caption{\small  Access ratio per level (Memtable, L0, L1+) and average read latency for each system.
  }
  \label{fig:eval_rrwr_read_ratio}
\end{figure}
\endgroup

\begin{figure*}[t]
  \centering
  \captionsetup[subfigure]{labelformat=empty}
  \subfloat[]{%
    \includegraphics[width=0.5\linewidth]{plots/eval_system_compair_legend.pdf}}
  \vspace{-20pt}
  \\[0pt]

  \hspace{-10pt}
  \includegraphics[width=0.165\linewidth]{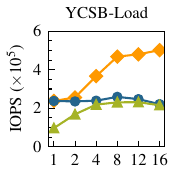}
  \hspace{-8pt}
  \includegraphics[width=0.14\linewidth]{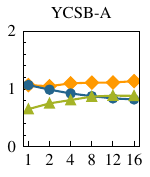}
  \hspace{-8pt}
  \includegraphics[width=0.14\linewidth]{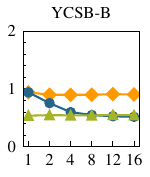}
  \hspace{-8pt}
  \includegraphics[width=0.14\linewidth]{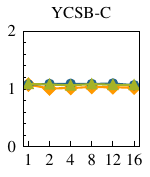}
  \hspace{-8pt}
  \includegraphics[width=0.14\linewidth]{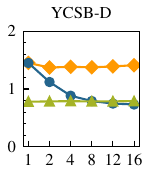}
  \hspace{-8pt}
  \includegraphics[width=0.16\linewidth]{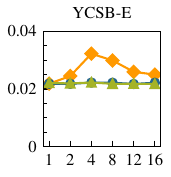}
  \hspace{-8pt}
  \includegraphics[width=0.152\linewidth]
  {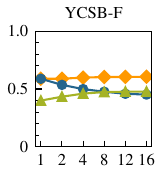}
  \hspace{-10pt}

  \vspace{-4pt}
  {\small\makebox[\linewidth][c]{\footnotesize Shard (\#) or Sub Comp. Thread (\#)}}
\vspace{-8pt}

  \vspace{-4pt}
  \caption{\small Comparison of throughput in the YCSB benchmark workload.}
  \vspace{-20pt}
  \label{fig:eval_ycsb}
\end{figure*}

Figure~\ref{fig:eval_rrwr_system_compair}(a) shows results for R90/W10. While ADOC remains nearly flat and RocksDB degrades sharply, \tbdfixed{} experiences only a modest decline. This difference stems from Memtable dynamics: with more shards (in \tbdfixed{}) or more Sub-Compaction threads (in RocksDB), the effective Memtable size per pipeline decreases, leading to faster flush cycles and fewer Memtable hits on reads. Although both systems use the same Memtable size overall, \tbdfixed{} maintains higher throughput than RocksDB because sharding prevents excessively frequent flushes per shard. A deeper explanation of this effect is given in Figure~\ref{fig:eval_rrwr_read_ratio}.

At R50/W50 (Figure~\ref{fig:eval_rrwr_system_compair}(b)), ADOC shows slight improvement, as the higher write ratio lets it to benefit from expanded Memtable capacity when more Sub-Compaction threads are enabled. 
Finally, at R10/W90 (Figure~\ref{fig:eval_rrwr_system_compair}(c)), the behavior closely resembles the write-only case in Figure~\ref{fig:eval_fr_system_compair}(b), where stalls dominate and \tbdfixed{}  outperforms the baselines.

\begingroup
\setlength{\textfloatsep}{0pt}
\begin{figure}[t]
  \centering
  \captionsetup[subfigure]{labelformat=empty}
  \subfloat[]{%
    \includegraphics[width=1\linewidth]{plots/eval_system_compair_legend.pdf}}
  \vspace{-20pt}
  \\[0pt]

  \hspace{-3pt}
  \includegraphics[width=0.35\linewidth]{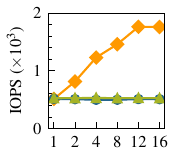}
  \hspace{-7pt}
  \includegraphics[width=0.325\linewidth]{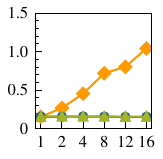}
  \hspace{-7pt}
  \includegraphics[width=0.325\linewidth]{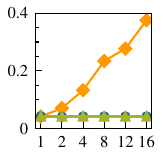}

  \vspace{-4pt}
  {\small\makebox[\linewidth][c]{\footnotesize Shard (\#) or Sub Comp. Thread (\#)}}
\vspace{-8pt}
  
  \hspace{-3pt}
  \makebox[0.35\linewidth]{\small{(a)  Length=1,024}}
  \hspace{-7pt}
  \makebox[0.325\linewidth]{\small{(b)  Length=4,096}}
  \hspace{-7pt}
  \makebox[0.325\linewidth]{\small{(c) Length=16,384}}

  \vspace{-5pt}
  \caption{\small  Comparison of throughput under different scan lengths in the YCSB-E workload.}
  \vspace{0pt}
  \label{fig:eval_iterator}
\end{figure}
\endgroup

Figure~\ref{fig:eval_rrwr_read_ratio} shows the average read latency and level-wise access ratios under the RRWR R50/W50 workload. In RocksDB (Figure~\ref{fig:eval_rrwr_read_ratio}(a)), increasing the number of Sub-Compaction threads leads to higher latency. This occurs because a larger number of threads reduces the Memtable size, accelerating flush cycles and lowering the probability of Memtable hits.

A similar trend is observed in \tbdfixed{} (Figure~\ref{fig:eval_rrwr_read_ratio}(b)), where Memtable hits decrease as the shard count grows. However, unlike RocksDB, sharding reduces the write load per shard, preventing flush cycles from becoming excessively frequent even with smaller Memtables. As a result, \tbdfixed{} sustains a higher Memtable hit ratio and effectively suppresses read-latency degradation.

ADOC (Figure~\ref{fig:eval_rrwr_read_ratio}(c)) exhibits the opposite trend. As the number of Sub-Compaction threads increases, read latency decreases. This is because faster L0 compaction reduces accesses to unsorted L0 SSTables, shifting queries toward more efficient hits in deeper, sorted levels.

\begin{table}[b]
    \centering 
    \caption{\small Description of YCSB workload characteristics.}
    \resizebox{\columnwidth}{!}{
    \tiny{
\begin{tabular}{c|ccc}
\toprule
\textbf{Type} & \textbf{Description} & \textbf{Distribution}  \\ \midrule
A & 50\% Updates, 50\% Reads & Zipfian  \\ 
B & 95\% Reads, 5\% Updates   & Zipfian  \\ 
C & 100\% Reads & Zipfian  \\ 
D & 95\% Reads, 5\% Inserts & Latest  \\ 
E & 95\% Seeks, 5\% Inserts; Scan length:128 & Uniform  \\ 
F & 50\% Read-Modify-Write, 50\% Reads & Zipfian  \\ \bottomrule
\end{tabular}
    }
    }
    \label{tbl:workload_ycsb}
\end{table}

\subsection{YCSB Benchmark Evaluation}
\label{eval:perf_ycsb}
Figure~\ref{fig:eval_ycsb} presents throughput comparisons across YCSB workloads, with workload details given in Table~\ref{tbl:workload_ycsb}.
Overall, \tbdfixed{} consistently achieves higher throughput than ADOC and RocksDB, regardless of the degree of parallelism.

One notable case is YCSB-E, which involves range queries. While ADOC and RocksDB remain largely unaffected by increasing parallelism, \tbdfixed{} shows throughput gains up to 4 shards, reaching about 1.7$\times$ higher performance than both baselines. Beyond this point, throughput declines as the shard count grows, since the relatively short scan length (128) amplifies the overhead of parallel range query execution.

Another observation is from YCSB-C, the read-only workload, where throughput remains unchanged across all systems. In this case, \tbdfixed{}, ADOC, and RocksDB all deliver comparable performance, as no write stalls are present to differentiate them.

\subsubsection{Iterator Efficiency}
\label{eval:perf_iterator}
Figure~\ref{fig:eval_iterator}(a)--(c) presents throughput results for the YCSB-E with scan lengths of 1K, 4K, and 16K. As the scan length increases, the benefit of parallelizing range queries outweighs the parallelization overhead introduced by larger shard counts, resulting in overall throughput improvements. Notably, \tbdfixed{} achieves $3.3\times$, $5.7\times$, and $8.9\times$ higher throughput than ADOC and RocksDB at scan lengths of 1K, 4K, and 16K, respectively.

\subsection{Impact of \tbdfixed{} Features}
\label{sec:eval_feature}

\begingroup
\setlength{\textfloatsep}{-10pt}
\begin{figure}[t]
  \centering

  \captionsetup[subfigure]{labelformat=empty}
  \subfloat[]{%
    \includegraphics[width=1\linewidth]{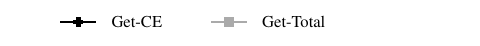}}
  \vspace{-20pt}
  \\[0pt]
  
\adjustbox{valign=t}
{\includegraphics[width=0.36\linewidth]{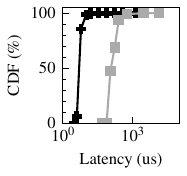}}
\hspace{-7pt}
\adjustbox{valign=t}{\includegraphics[width=0.31\linewidth]{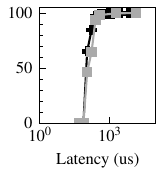}}
\hspace{3pt}
\adjustbox{valign=t}{%
  \raisebox{-6ex}{ 
  {\footnotesize
  \begin{tabular}{c|c}
   \toprule
    \textbf{Cap} & \textbf{IOPS} \\
    \midrule
    \textbf{On} & 6.58K \\ 
    \textbf{Off} & 6.35K \\
    \bottomrule
  \end{tabular}
  }}
}

  \makebox[0.34\linewidth]{\small{(a) CE Capping On}}
  \makebox[0.34\linewidth]{\small{(b) CE Capping Off}}
  \hspace{-4pt}
  \makebox[0.24\linewidth]{\small{(c) Throughput}}

  \vspace{-5pt}
  \caption{\small Latency and throughput comparison under configurations with and without CE Capping. Get-CE is the time to read a CE, and Get-Total is the overall time to process a Get.
  }
  \label{fig:eval_ce}
\end{figure}
\endgroup

\subsubsection{CE L0 Capping Efficiency}
\label{eval:feature_ce}
{Although the preceding experiments were performed with CE Capping enabled, Figures~\ref{fig:eval_ce}(a)--(b) present a comparison of read latency in \tbdfixed{} (16) with and without CE Capping, under the db\_bench ReadRandom (100\% read) workload following FillRandom with 8KB values.}
When CE Capping was disabled, the CE access time was almost the same as the overall read latency. However, when Capping was enabled, the CE read time decreased by 95.5\%, resulting in a noticeable improvement in latency.
Figure~\ref{fig:eval_ce}(c) shows that throughput improvement depends on how many CE reads occur. In this workload, CE reads were only about 15\%, so the gain was small. If the ratio is higher, the reduced latency leads to larger throughput gains, making CE Capping more effective.

\begin{figure}[t]
  \centering

  \hspace{-6pt}
  \includegraphics[width=0.49\linewidth]{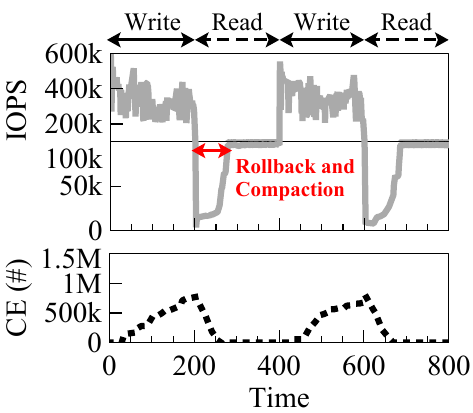}
  \hspace{0pt}
  \includegraphics[width=0.49\linewidth]{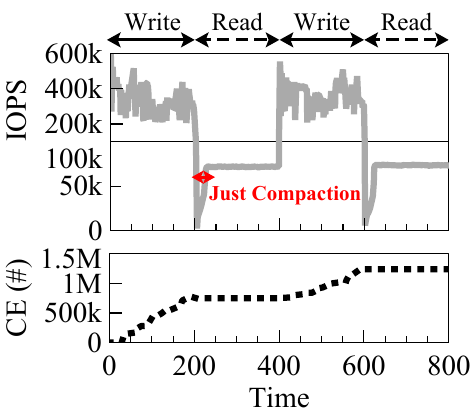}

  \vspace{-2pt}
  
  \hspace{-6pt}
  \makebox[0.49\linewidth]{\small{(a) Rollback On}}
  \hspace{0pt}
  \makebox[0.49\linewidth]{\small{(b) Rollback Off}}

  \vspace{-4pt}
  \caption{\small  Overhead and throughput comparison under rollback execution.
  }
  \vspace{-16pt}
  \label{fig:eval_rollback}
\end{figure}

\begin{figure}[t]
  \centering

  \captionsetup[subfigure]{labelformat=empty}
  \subfloat[]{
    \includegraphics[width=1\linewidth]{plots/eval_system_compair_legend.pdf}}
  \vspace{-20pt}
  \\[0pt]

  \hspace{-3pt}
  \includegraphics[width=0.36\linewidth]{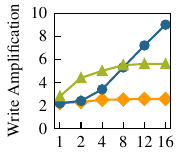}
  \hspace{-7pt}
  \includegraphics[width=0.31\linewidth]{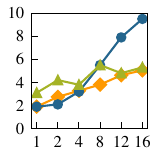}
  \hspace{-7pt}
  \includegraphics[width=0.31\linewidth]{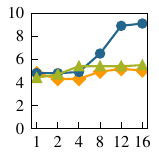}

  \vspace{-4pt}
  {\small\makebox[\linewidth][c]{\footnotesize Shard (\#) or Sub Comp. Thread (\#)}}
\vspace{-8pt}
  
  \hspace{-3pt}
  \makebox[0.36\linewidth]{\small{(a) 128B}}
  \hspace{-7pt}
  \makebox[0.31\linewidth]{\small{(b) 1KB}}
  \hspace{-7pt}
  \makebox[0.335\linewidth]{\small{(c) 8KB}}

  \vspace{-4pt}
  \caption{\small  Write amplification comparison for each system.
  }
  \vspace{0pt}
  \label{fig:eval_wa}
\end{figure}

\subsubsection{Rollback Overhead and Throughput Benefits}
\label{eval:feature_rollback}
Figures~\ref{fig:eval_rollback}(a)--(b) present throughput and CE counts in \tbdfixed{} (16) while alternately running 8KB FillRandom and ReadRandom workloads for 200 seconds each, comparing rollback enabled and disabled. In the initial write phase, approximately 850,000 CEs are generated. During the subsequent read phase, the system performs compaction prior to serving requests, and enabling rollback incurs about 60 seconds of additional delay. Despite this overhead, enabling rollback delivers throughput improvement over the disabled case, demonstrating that rollback effectively enhances read performance.
As described in Section~\ref{sec:design_internal}, rollback is a user-defined policy that benefits read-heavy workloads. It is not just overhead but can improve throughput. Since workload-aware decision making has been well studied, we leave adaptive rollback policies for future work.
{Note that all experiments except for this evaluation were conducted with rollback disabled to maintain consistency.} 

\subsubsection{Write and Space Amplification}
\label{eval:feature_ampl}
Figure~\ref{fig:eval_wa} presents the comparison of Write Amplification (WA) across value sizes under the FillRandom workload. \tbdfixed{} exhibited more stable WA than other systems, indicating that CE insertion has little impact on WA. In contrast, ADOC and RocksDB show increasing WA as the number of Sub-Compaction threads grows, since SSTable fragmentation increases. In particular, for RocksDB, as Memtable size decreases and the number of Memtables increases, the number of L0 SSTable files grows sharply, causing WA to deteriorate significantly.

\begin{table}[b]
    \centering 
    \caption{\small Space amplification in {\tbdfixed{}}.
    }
    \renewcommand{\arraystretch}{1.2}
    \vspace{-6pt}
    \resizebox{\columnwidth}{!}{
    \scriptsize{
\begin{tabular}{c|cccccc}
\toprule
\textbf{LSM shards (\#)} & 1 & 2 & 4 & 8 & 12 & 16 \\ \midrule
\textbf{Space Amplification} & 1 & 1.28 & 1.35 & 1.5 & 1.21 & 1.14 \\ 
\textbf{CE (\#)} & - & 3.5M & 3.4M & 5.3M & 2M & 1.5M \\ \bottomrule
\end{tabular}
    }
    }
    \label{tbl:eval_amp_sa}
\end{table}

\noindent\textbf{Space Amplification.}
Table~\ref{tbl:eval_amp_sa} presents the number of CEs and the Space Amplification (SA) of \tbdfixed{} in the FillRandom 8KB workload(Figure~\ref{fig:eval_fr_system_compair}(c)) across different shard counts. 
As the number of shards increases up to 8, the likelihood of avoiding write stalls rises, resulting in more CEs and higher SA.  Beyond 12 shards, however, the sharding effect becomes more pronounced, reducing per-shard write stalls and thereby lowering both CE occurrences and SA. These results indicate that increasing the shard count enhances not only throughput but also space efficiency.

\begin{figure}[t]
  \centering

  \captionsetup[subfigure]{labelformat=empty}
  \subfloat[]{
    \includegraphics[width=1\linewidth]{plots/eval_system_compair_legend.pdf}}
  \vspace{-20pt}
  \\[0pt]

  \hspace{-3pt}
  \includegraphics[width=0.35\linewidth]{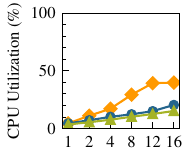}
  \includegraphics[width=0.33\linewidth]{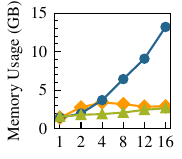}

  \vspace{-4pt}
  {\small\makebox[\linewidth][c]{\footnotesize Shard (\#) or Sub Comp. Thread (\#)}}
\vspace{-8pt}
  
  \hspace{-3pt}
  \makebox[0.35\linewidth]{\small{(a) CPU}}
  \makebox[0.33\linewidth]{\small{(b) Memory}}

  \vspace{-4pt}
  \caption{\small  Resource usage comparison for each system.
  }
  \vspace{-12pt}
  \label{fig:eval_resource}
\end{figure}

\label{eval:feature_resource}
\begin{figure}[t]
  \centering

  \captionsetup[subfigure]{labelformat=empty}
  \subfloat[]{%
    \includegraphics[width=1\linewidth]{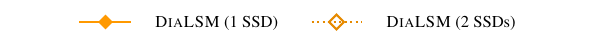}}
  \vspace{-20pt}
  \\[0pt]

  \hspace{-3pt}
  \includegraphics[width=0.35\linewidth]{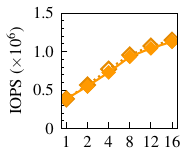}
  \includegraphics[width=0.35\linewidth]{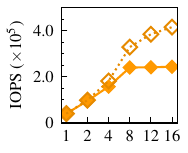}

  \vspace{-4pt}
  {\small\makebox[\linewidth][c]{\footnotesize Shard (\#) or Sub Comp. Thread (\#)}}
\vspace{-8pt}

  \hspace{-3pt}
  \makebox[0.35\linewidth]{\small{(a) 128B}}
  \makebox[0.35\linewidth]{\small{(b) 8KB}}

  \vspace{-4pt}
  \caption{\small Evaluation of \tbdfixed{} utilizing multiple devices.
  }
  \vspace{-4pt}
  \label{fig:eval_multidevice}
\end{figure}

\subsubsection{Resource Utilization}
Figures~\ref{fig:eval_resource}(a)--(b) show CPU and memory consumption under the FillRandom 8KB.
\tbdfixed{} uses more CPU to accelerate parallel compaction, reducing total processing time. Its CE caching adds some memory overhead, but modestly. In contrast, RocksDB’s memory usage rises sharply as immutable Memtables accumulate, showing that \tbdfixed{} achieves higher memory efficiency even with smaller, more frequent Memtables.

\subsection{Multi-SSD Performance Evaluation}

In large-scale services (e.g., ByteDance~\cite{bytedance}), multiple SSDs are commonly deployed to meet throughput and parallelism demands, with RocksDB serving as the backend storage engine. While a single SSD often suffers from controller bottlenecks, scaling out with multiple SSDs alleviates these limitations and delivers substantial performance gains. Similarly, \tbdfixed{} mitigates LSM pipeline bottlenecks inherent to monolithic LSM tree designs through sharding, but device-level bottlenecks still remain. 
To address this, distributing LSM shards across multiple SSDs offers a practical solution. We experimentally validate this by comparing single-SSD and dual-SSD setups, placing LSM shards evenly across the two devices and running FillRandom workloads. 
As shown in Figure~\ref{fig:eval_multidevice}, performance remains comparable for 128B values, but with 8KB workloads, a single SSD exhibits clear bottlenecks once the shard count exceeds eight. 
Beyond this point, distributing LSM shards across dual SSDs achieves performance scaling proportional to the available bandwidth. 
These results highlight the importance of intelligent shard placement in multi-SSD arrays, suggesting that in multi-tenant cloud environments, effective load balancing of LSM shards across SSD controllers is a critical challenge.

\begingroup
\setlength{\textfloatsep}{0pt}
\begin{figure}[t]
  \centering

  \captionsetup[subfigure]{labelformat=empty}
  \subfloat[]{%
    \includegraphics[width=1\linewidth]{plots/eval_system_compair_legend.pdf}}
  \vspace{-20pt}
  \\[0pt]

  \hspace{-5pt}
  \includegraphics[width=0.33\linewidth]{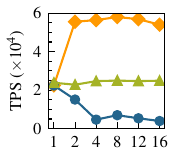}
  \hspace{-7pt}
  \includegraphics[width=0.33\linewidth]{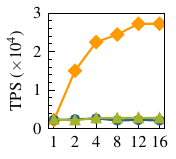}
  \hspace{-7pt}
  \includegraphics[width=0.35\linewidth]{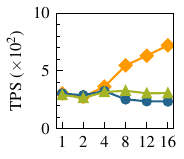}
  \hspace{-5pt}

  \vspace{-4pt}
  {\small\makebox[\linewidth][c]{\footnotesize Shard (\#) or Sub Comp. Thread (\#)}}
\vspace{-10pt}
  
  \hspace{-5pt}
  \makebox[0.33\linewidth]{\small{(a) oltp\_insert}}
  \hspace{-7pt}
  \makebox[0.33\linewidth]{\small{(b) oltp\_write\_only}}
  \hspace{-7pt}
  \makebox[0.35\linewidth]{\small{(c) oltp\_read\_write}}
  \hspace{-5pt}

  \vspace{-2pt}
  \caption{\small Sysbench evaluation of MariaDB with MyRocks.
  }
  \vspace{-2pt}
  \label{fig:eval_rdbms}
\end{figure}
\endgroup

\subsection{OLTP Evaluation on RDBMS}
\label{eval:rdmbs}
We evaluate effectiveness in RDBMS environments using Sysbench OLTP workloads on MariaDB based on RocksDB, known as MyRocks~\cite{matsunobu2020myrocks}. Detailed characteristics of the OLTP workloads are presented in Table~\ref{tbl:workload_oltp}, and the workloads are executed with 10 I/O threads for 1800 seconds.

Figure~\ref{fig:eval_rdbms}(a) shows the results for oltp\_insert, which consists of 100\% INSERT operations. Starting from two shards, write stalls are alleviated, achieving up to 2.3$\times$ higher throughput than ADOC, after which performance saturates. In contrast, RocksDB exhibits degraded performance due to reduced Memtable sizes, similar to the behavior observed from eight shards in Figure~\ref{fig:eval_fr_system_compair}(c).
Figure~\ref{fig:eval_rdbms}(b) shows the results for oltp\_write\_only, which includes DELETE and UPDATE operations in addition to INSERT. RocksDB and ADOC show limited benefits from Sub-Compaction, whereas \tbdfixed{} fully exploits sharding effects and achieves up to 9.9$\times$ higher throughput.
Figure~\ref{fig:eval_rdbms}(c) shows the results for oltp\_read\_write. While \tbdfixed{} achieves 2.2$\times$ higher throughput due to sharding effects, the other systems exhibit little change or a slight decrease in performance.

These results indicate that \tbdfixed{}'s design can be effectively applied beyond key-value stores to RDBMSs.

\begingroup
\begin{table}[h]
    \centering 
    \vspace{-2pt}
    \caption{\small Description of OLTP workload characteristics.}
    \vspace{-4pt}
    \resizebox{\columnwidth}{!}{
    \tiny{
\begin{tabular}{c|c}
\toprule
\textbf{Workload Type} & \textbf{Operations per Transaction (Execution Order)} \\ \midrule
{oltp\_insert} & 1 \texttt{INSERT} \\ 
{oltp\_write\_only} & 2 \texttt{UPDATE}, 1 \texttt{DELETE}, 1 \texttt{INSERT} \\ 
{oltp\_read\_write} & 10 \texttt{SELECT}, 2 \texttt{UPDATE}, 1 \texttt{INSERT}, 1 \texttt{DELETE} \\ \bottomrule
\end{tabular}
    }
    }
    \label{tbl:workload_oltp}
\end{table}
\vspace{-10pt}
\endgroup
\section{Related Work}
\noindent\textbf{Optimizing the Monolithic Pipeline.}
Extensive work accelerates compaction and improves read/write amplification trade-offs in a single LSM-tree. Software approaches improve compaction scheduling~\cite{zhang2014pipelined,balmau2019silk, balmau2020silk+,liang2021cruisedb,hu2023aislsm,pan2017dcompaction,byun2025revisiting}, tune system parameters~\cite{yu2023adoc,huynh2021endure,wang2025rethinking,thakkar2024llmtune,thakkar2026tellytune}, modify compaction strategies~\cite{dayan2022spooky,wang2022reducing,dayan2017monkey,dayan2018dostoevsky} (e.g., leveled and tiered), redesign data layouts~\cite{teng2017lsbm,lepers2019kvell,chen2021spandb,raina2023efficient,xanthakis2024vlsm,wu2015lsm,balmau2017triad,ren2017slimdb,sun2023improving,raju2017pebblesdb,zhao2021wipdb,ting2019geardb}, reduce syscall overhead~\cite{byun2026resystance}, or separate keys from values~\cite{lu2017wisckey,blobdb,intergrated_blobdb,li2021differentiated,jamil2025dedupkv}.

Hardware-assisted methods offload compaction to FPGAs~\cite{zhang2020fpga,sun2020fpga,tang2024stem}, DPUs~\cite{ding2023dcomp,ding2024d2comp,chen2023iknowfirst}, GPUs~\cite{sun2025gparakv,sun2024glsm,sun2025rgkv,zhou2024gpu,zhou2025gpcomp,xu2020luda}, KV-SSDs~\cite{kim2025kvaccel}, or distributed and cloud backends~\cite{wang2023mirrorkv,yu2024caas,kim2024coordinating,kim2025eco,lin2026o3}, and exploit in-storage processing~\cite{sun2025a,sun2019collaborative} or non-volatile memory~\cite{yao2020matrixkv,ding2022trianglekv,zhang2021nvlsm,chen2023workload} such as 3D XPoint~\cite{hady2017platform}; others systematically analyze compaction designs and future directions~\cite{sarkar2022constructing,sarkar2022compactionary}.
Beyond compaction-focused optimizations, a wide range of studies have actively explored optimizing the overall structure and operation of LSM-trees~\cite{zhu2025mnemosyne,mo2023learning,sarkar2020lethe,sarkar2022dissecting,luo2018efficient,zhong2025disco,lee2023iterator}.

\noindent\textbf{Partitioning and Multi-Tree Approaches.}
Decomposing a monolithic index into smaller, independently maintained units is well established, from partitioned and multi-version B-trees~\cite{graefe2003sorting,graefe2012concurrency,riegger2017write,riegger2020mv} to fragmented or range-partitioned LSM-trees such as PebblesDB~\cite{raju2017pebblesdb} and WipDB~\cite{zhao2021wipdb}; the latter compact fragments independently but can trap hot ranges in continuous merges, while Column Families~\cite{columnfamiles} separate trees only logically and require explicit user management. We therefore do not claim the decomposition principle itself as novel: \tbdfixed{}'s contribution is the LSM-internal stall mitigation built on hash sharding---dynamic fallback rerouting, CE indirection, and rollback/recovery integration---which sustains writes under stalls and lowers stall probability with shard count, achieving near-stall-free operation in software.
\vspace{-8pt}
\section{Discussion}
\label{sec:discuss}
\noindent\textbf{Why Internal Sharding is Necessary.} System-level sharding, as in ZippyDB~\cite{zippydb} or Kvrocks~\cite{kvrocks}, improves scalability but does not eliminate stalls within each embedded store. In contrast, internal sharding allows finer-grained control, better use of device-level parallelism (e.g., NVMe SSDs), and avoids the overhead of inter-node communication. While it introduces modest  CPU Utilization, the gains in latency and stability far outweigh these costs (§\ref{sec:eval}).

\noindent\textbf{Number of Shards.}
Determining an appropriate shard count is inherently challenging: CPU, memory, I/O bandwidth, workload skew, and the deployment environment jointly determine how sharding affects write stalls, so no single setting is universally optimal. Dynamic adjustment is feasible but incurs nontrivial migration and rebalancing overhead; we therefore treat shard count as a deployment-layer concern (as in systems such as HDFS) rather than an internal LSM optimization.

\vspace{-8pt}
\section{Conclusions}
\label{sec:conc}
We showed that write stalls in monolithic LSM trees are a structural limitation of the tightly coupled write–flush–compaction pipeline and cannot be eliminated by parameter tuning alone. Based on a vacation-based queuing analysis, we proposed \tbdfixed{}, an internally shard-based LSM architecture that substantially reduces write stalls and achieves up to 2.4$\times$ higher throughput with lower latency than state-of-the-art approaches. 

\bibliographystyle{ieeetr}
\bibliography{ref}

\end{document}